\documentclass{aa}
\usepackage[varg]{txfonts}
\usepackage{graphicx}
\usepackage{amsmath}
\usepackage{natbib}
\usepackage{color}
\usepackage[usenames,dvipsnames]{xcolor}
\usepackage{float}
\usepackage{fixltx2e}
\usepackage{caption,subcaption}
\usepackage[breaklinks, colorlinks, citecolor=blue]{hyperref}
\usepackage{tabularx}
\usepackage[export]{adjustbox}

\bibpunct{(}{)}{;}{a}{}{,}

\newcommand{\Om}{\Omega_\textrm{m}}
\newcommand{\sig}{\sigma_\textrm{8}}
\newcommand{\Ho}{H_\textrm{0}}
\newcommand{\Ob}{\Omega_\textrm{b}}
\newcommand{\Tcmb}{T_\textrm{CMB}}
\newcommand{\ns}{n_\textrm{s}}
\newcommand{\neff}{N_\textrm{eff}}
\newcommand{\wo}{w_\textrm{0}}
\newcommand{\wa}{w_\textrm{a}}
\newcommand{\smnu}{\sum{m_\nu}}
\newcommand{\Planck}{{\em Planck}}
\newcommand{\Euclid}{{\em Euclid}}
\newcommand{\RomanT}{{\em Roman}}
\newcommand{\Rubin}{{\em Rubin}}

\begin{document}

\title{Impact of the calibration of the Halo Mass Function  on galaxy cluster number count cosmology} 

\subtitle{} 

\author{Emmanuel Artis\inst{1}
\and Jean-Baptiste Melin\inst{1}
  \and James G. Bartlett\inst{2,3}
   \and Calum Murray\inst{2}} 

\offprints{----------, \email{----------}}

\institute{IRFU, CEA, Universit\'e Paris-Saclay, F-91191 Gif-sur-Yvette, France
  \and Universit\'e de Paris, CNRS, Astroparticule et Cosmologie,  F-75006 Paris, France
  \and Jet Propulsion Laboratory, California Institute of Technology, Pasadena, California, USA} 

\date{Received ----------/ Accepted ------------}

\abstract {The halo mass function (HMF) is a critical element in cosmological analyses of galaxy cluster catalogs.  We quantify the impact of uncertainties in HMF parameters on cosmological constraints from cluster catalogs similar to those from \Planck, those expected from the \Euclid, \RomanT\ and Rubin surveys, and from a hypothetical larger future survey.   We analyse simulated catalogs in each case, gradually loosening priors on HMF parameters to evaluate the degradation in cosmological constraints.  While current uncertainties on HMF parameters do not substantially impact \Planck-like surveys, we find that they can significantly degrade the cosmological constraints for a \Euclid-like survey.  Consequently, the current precision on the HMF will not be sufficient for \Euclid\ (or \RomanT\ or Rubin) and possible larger surveys.  Future experiments will have to properly account for  uncertainties in HMF parameters, and it will be necessary to improve precision of HMF fits to avoid weakening constraints on cosmological parameters.}

\maketitle

\section{Introduction}
\label{sec:intro}
Galaxy clusters are unique tools for  astrophysical and cosmological studies ranging from galaxy formation to fundamental physics \citep{Voit2005, 2011ARA&A..49..409A, KB2012}. 
In particular, it is now well established that their number density $n(M,z)$ is a powerful cosmological probe, notably of  the matter density, $\Om$, the root mean square density fluctuation, $\sig$, the dark energy equation-of-state and the sum of the neutrino masses \citep{Weinberg+2013}.  For this reason, cluster number counts figure among the four primary dark energy probes.

Surveys across different wavebands have used cluster number counts to derive cosmological constraints \citep[e.g.,][]{Vikhlinin+2009, Mantz+2010, Rozo+2010,Hasselfield+2013,  2016A&A...594A..24P, 2019ApJ...878...55B, 10.1093/mnras/stz1949, Costanzi+2020}. 
These analyses are based on the halo mass function (HMF hereafter) originating in the Press \& Schechter \citep{1974ApJ...187..425P} formalism. The HMF provides the key link between  cosmological parameters and the number density of clusters; it essentially bundles the complicated non-linear physics of halo formation into a simple analytical formula involving only the linear power spectrum and other linear-theory quantities.  This is essential because it avoids the running of costly N-body simulations at each point in parameter space when establishing cosmological constraints. 

Current cluster count analyses use catalogs of hundreds (in millimeter and X-ray data) to thousands (in optical data) of clusters, and they are not limited by our knowledge of the HMF. But uncertainties on the HMF could become important as future catalogs grow by factors of 10 to 100. \Euclid\ will detect tens of thousands of clusters \citep{2016MNRAS.459.1764S} as will  {\em eROSITA} \citep{2017A&A...606A.118H}, Simons Observatory \citep{Ade_2019}, \Rubin~\citep{Rubin} and CMB-S4 \citep{abazajian2019cmbs4}. In the longer term, {\em Roman}~\citep{2015JPhCS.610a2007G}, {\em PICO}~\citep{2019arXiv190807495H} and a Voyage 2050 {\em Backlight} mission~\citep{Basu2019} could detect from hundreds of thousands to one million clusters.

\begin{figure}%[H]
\includegraphics[width=9cm]{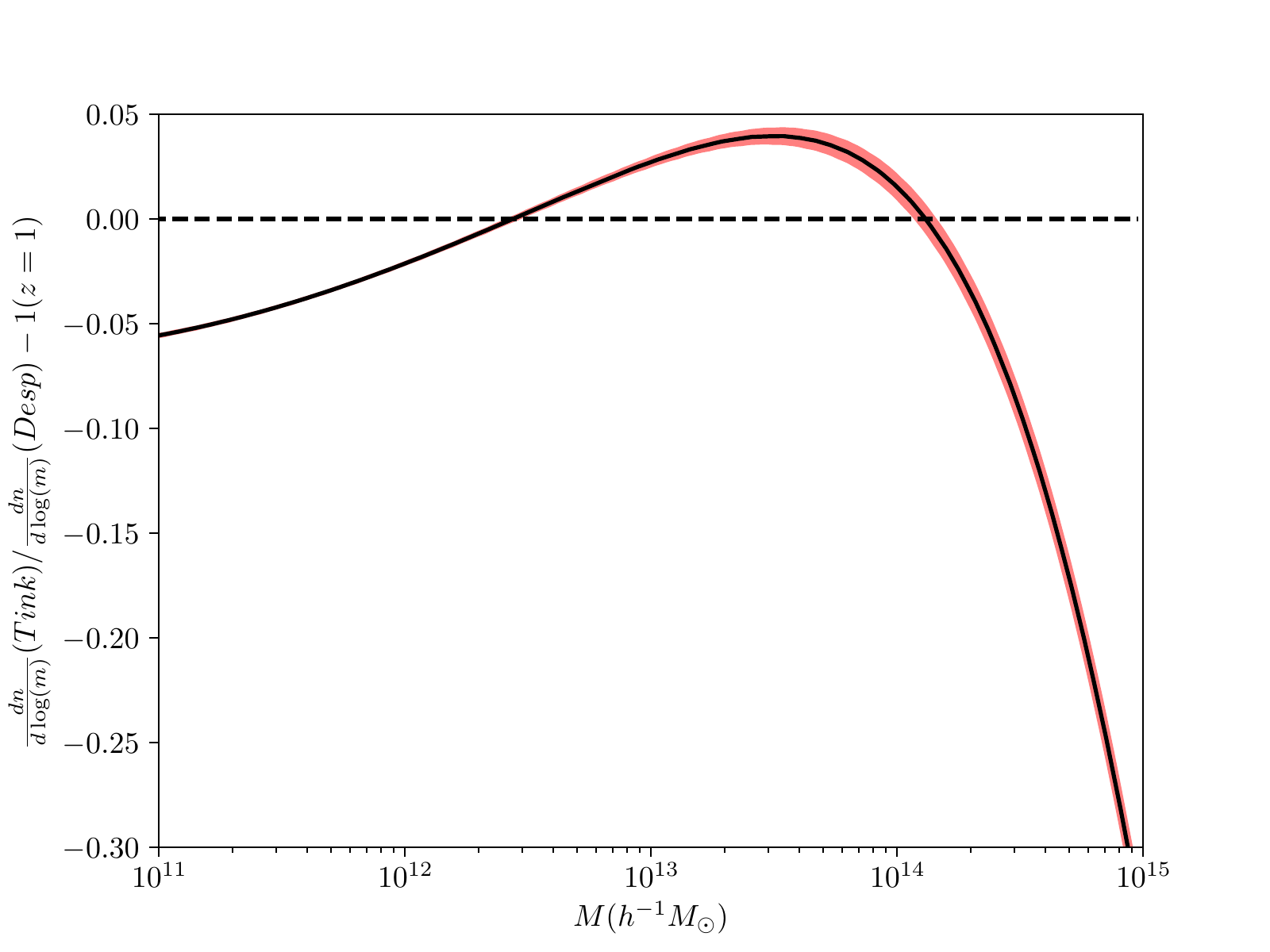}
\caption{Relative difference of the \cite{2008ApJ...688..709T} and the \cite{Despali_2015} mass functions at z=1 for a $\Lambda$CDM cosmology. In red, the convex envelope of all the scenarios covered by the error bars provided by~\cite{Despali_2015}. The mass is defined with respect to the virial radius.}
\label{comp_tinker_desp}
\end{figure}

As statistical errors on cluster counts  dramatically drop, effects now considered as second order will begin to noticeably contribute to the uncertainty budget of cosmological analyses. In this {\em paper}, we examine the potential importance of uncertainties related to modeling errors in the HMF. Popular HMF formulas like \cite{2008ApJ...688..709T} predict cluster counts in $\Lambda$CDM to a precision of several percent ($\sim5$ to 10\%) (compared to full N-body simulations). The recent \cite{Despali_2015} HMF provides an expression with parameters and associated errors fitted on $\Lambda$CDM N-body simulations. Fig.~\ref{comp_tinker_desp} presents the relative difference of the Tinker and Despali mass functions at z=1 as a function of the virial mass. The difference is below $5\%$ for masses less than a few $10^{14} \, M_\odot$, so within the quoted error bars for the Tinker mass function. The red band delineates the uncertainty from the HMF parameters provided by~\cite{Despali_2015}. They are much smaller than the difference between the two mass functions, bringing to light the difference between the {\em precision} of a given HMF formula and its potential bias (a question of {\em accuracy}). The origin of this difference is beyond the scope of our study.  

Instead, we take the Despali mass function and its given uncertainty, shown by the red band, as representative of  our knowledge of the mass function. This is clearly optimistic faced with the importance of the difference between the Tinker and other mass functions. Nevertheless, we will see that even in the absence of any bias, these uncertainties may have to be accounted for in future cosmological analyses of large galaxy cluster surveys.

Different approaches can be chosen to model the uncertainties related to the HMF~\citep[see e.g.][]{10.1093/mnras/stz1949}. We investigate them by allowing the parameters of the HMF to vary according to priors set by its fit to simulations, such as provided by Despali et al. By modifying the size of these priors, we forecast the level of precision required to avoid significant degradation of cosmological parameter estimation. We require that the HMF parameters be known to sufficient precision so that:
\begin{itemize}
    \item The precision on cosmological parameters is degraded by less than 5\%, and
    \item The precision on the Figure-of-Merit (FoM, the inverse of the area of the confidence ellipse) is degraded by less than 10\%.
\end{itemize}

We emphasize that this approach assumes that the fitted HMF yields unbiased cosmological constraints. The uncertainties on the HMF parameters  depend primarily on the number of halos in the simulation used to fit (or "calibrate") the HMF, a number that potentially far exceeds the number of clusters in the universe. But we see from the example of Figure \ref{comp_tinker_desp} that the difference between the HMF forms from \cite{2008ApJ...688..709T} and \cite{Despali_2015} largely exceeds the statistical uncertainties shown by the red band.  The statistical uncertainties do not account for the accuracy, or potential bias, of the HMF fitting formula.  We will however consider only the statistical uncertainties in the present study, returning briefly to the issue of bias in Sect. \ref{discussion}.

The article is organized as follows. In Sect.~\ref{sec:methods}, we give a brief overview of the mass function, describe our likelihood, explain how we compute our Fisher matrices, and finally list the cosmological models and cluster catalogs that we examine. We present results in Sect.~\ref{sec:results} for each cosmological model, and then summarize in Sect.~\ref{sec:summary} before a concluding discussion in Sect.~\ref{discussion}.\\

\section{Methods}
\label{sec:methods}

\subsection{Mass function and cluster likelihood}
The mass function formalism was originally introduced by Press and Schechter \citep{1974ApJ...187..425P}.  It predicts the comoving number density of halos, $n(M,z)$, for a given set of cosmological parameters as
\begin{equation}
\label{PS}
\frac{\textrm{d}n}{\textrm{d}\ln M}(M,z)=\frac{\bar{\rho}}{M}\left | \frac{\textrm{d} \ln(\sigma(M,z))}{\textrm{d}\ln M} \right | f(\sigma(M,z))
\end{equation}
where
$\bar{\rho}$ is the mean density of the universe today, $f(\sigma)$ is the multiplicity function (discussed hereafter) and
\begin{equation}
\label{rms}
\sigma^2 (M,z)=\int{\frac{1}{2\pi^2}P_{\mathrm{lin}}(k,z)W^2 [kR(M)]k^2\mathrm{d}k},
\end{equation}
with $P_\mathrm{lin}(k,z)$ the {\em linear power spectrum} at redshift $z$, $R(M)=[M/(4\pi \bar{\rho}/3)]^{1/3}$ and $W(kR)=[3/(kR)^2][\sin(kR)/(kR)-\cos(kR)]$ is the spherical top-hat window function. Although the classical Press and Schechter approach  provides a good approximation under the assumption of a simple Gaussian process for $f(\sigma)$, numerous recent studies (based on numerical simulations or solving theoretical issues) have devised more accurate parametrizations of the multiplicity function  \citep[e.g.][]{1991ApJ...379..440B, Lacey_1994, Sheth1999, Jenkins2001, Crocce2010, PhysRevD.84.023009, Despali_2015}.\\

Relying on the Le SBARBINE simulations, the main interest of the Despali mass function (hereafter DMF), compared with other widely used HMF like Tinker \citep{2008ApJ...688..709T}, is the claim of universality, i.e., the parameters used to describe the multiplicity function $f(\sigma)$ do not explictly depend on either the redshift of the clusters or the cosmology. This is important because it means that when performing a cosmological analysis, the universal shape can be used at any redshift and in any cosmological scenario. This universality, however, requires use of $M_{\mathrm{vir}}$, the cluster mass defined with respect to the virial overdensity $\Delta_c$~\citep[e.g.][]{Lahav1991}. 

The shape of the function is the following, originally from \cite{Sheth1999}:
\begin{equation}
\label{desp}
\nu f(\nu)= A_0 \left(1+\frac{1}{\nu'^p}\right)\left(\frac{\nu'}{2\pi}\right)^{1/2}\mathrm{e}^{-\nu'/2},
\end{equation}
where $\nu'=a\nu$. The parameter $\nu$ is related to the root mean square density fluctuation via
\begin{equation}
\label{nu}
\nu = \delta_c^2 (z)/\sigma^2 (M,z),
\end{equation}
where $\delta_c(z)$ is the critical linear theory overdensity, which can be approximated by \cite{Kitayama1996}
\begin{equation}
\label{ov}
\delta_c (z) \approx \frac{3}{20} (12 \pi)^{2/3} [1+0.0123 \log(\Om(z))],
\end{equation} 
where $\log$ refers to logarithm in base 10.

We are then left with three parameters $(a,p,A_0)$ describing, respectively, the high-mass cutoff, the shape at lower masses, and the normalization of the curve. Once again, if the virial overdensity is used, no change in these parameters is expected with respect to either redshift or cosmology.

We compute the expected number density of objects per unit of mass, redshift, and area on the sky $\Omega$ as
\begin{equation}
\label{exp0}
\partial_{\small{{z,\ln{M},\Omega}}}N(M,z)=\frac{\bar{\rho}}{M}\left | \frac{\textrm{d} \ln(\sigma(M,z))}{\textrm{d}\ln M} \right | f(\sigma(M,z))\partial_{z,\Omega}V
\end{equation}
where $\partial_{z,\Omega}V$ is the comoving volume per unit redshift and solid angle at the considered redshift.
The quality of the cosmological parameter estimation with cluster number counts primarily depends on the number of clusters in the survey.
Since we focus on the effect of the HMF parameterization, we do not consider any observable-mass relation in this work. We also assume that we can directly measure  cluster mass and that the selection function can be modeled by a simple mass threshold $M_{\rm thres}$. We present the survey cases studied (defined by $M_{\rm thres}$, $\Omega$, redshift range $[z_\mathrm{min},z_\mathrm{max}]$) in Sect.~\ref{sec:studycases}.  %---------

For each survey-type, we estimate the cosmological parameters
$$\Theta=\{\Om,\sig,\wo,...\}$$
and simultaneously constrain the Despali HMF parameters, 
$$\Xi=\{a,p,A_0\}.$$
We only consider shot noise, so cosmic variance is neglected. Instead of using the \cite{1979ApJ...228..939C} statistics in bins of mass and redshift, we compute the unbinned Poisson likelihood of the two dimensional point process:
\begin{multline}
\label{cash}
\ln \mathcal{L}(\Theta,\Xi)=\quad -\quad \int_{M_\mathrm{thres}}^{+\infty} \int_{z_\mathrm{min}}^{z_\mathrm{max}}  \partial_{z,\ln{M}} N(M,z)\mathrm{d}\ln{M}\mathrm{d}z
\\+\quad\sum_{i=1}^{N_\mathrm{clusters}} \ln\left( \partial_{z,\ln{M}} N(M_i,z_i) \right) 
\end{multline}
where the sum runs over individual clusters in the catalog and $N_\mathrm{clusters}$ is the total number of clusters.
\subsection{Fisher matrix}
\label{sec:fisher}
The aim of our Fisher analysis is to quantify the precision on the HMF parameters, $\Xi$, required to limit the impact of their uncertainty on the cosmological constraints.
%-------------------------------------------------------
\begin{figure}
\includegraphics[width=10cm,height=8cm]{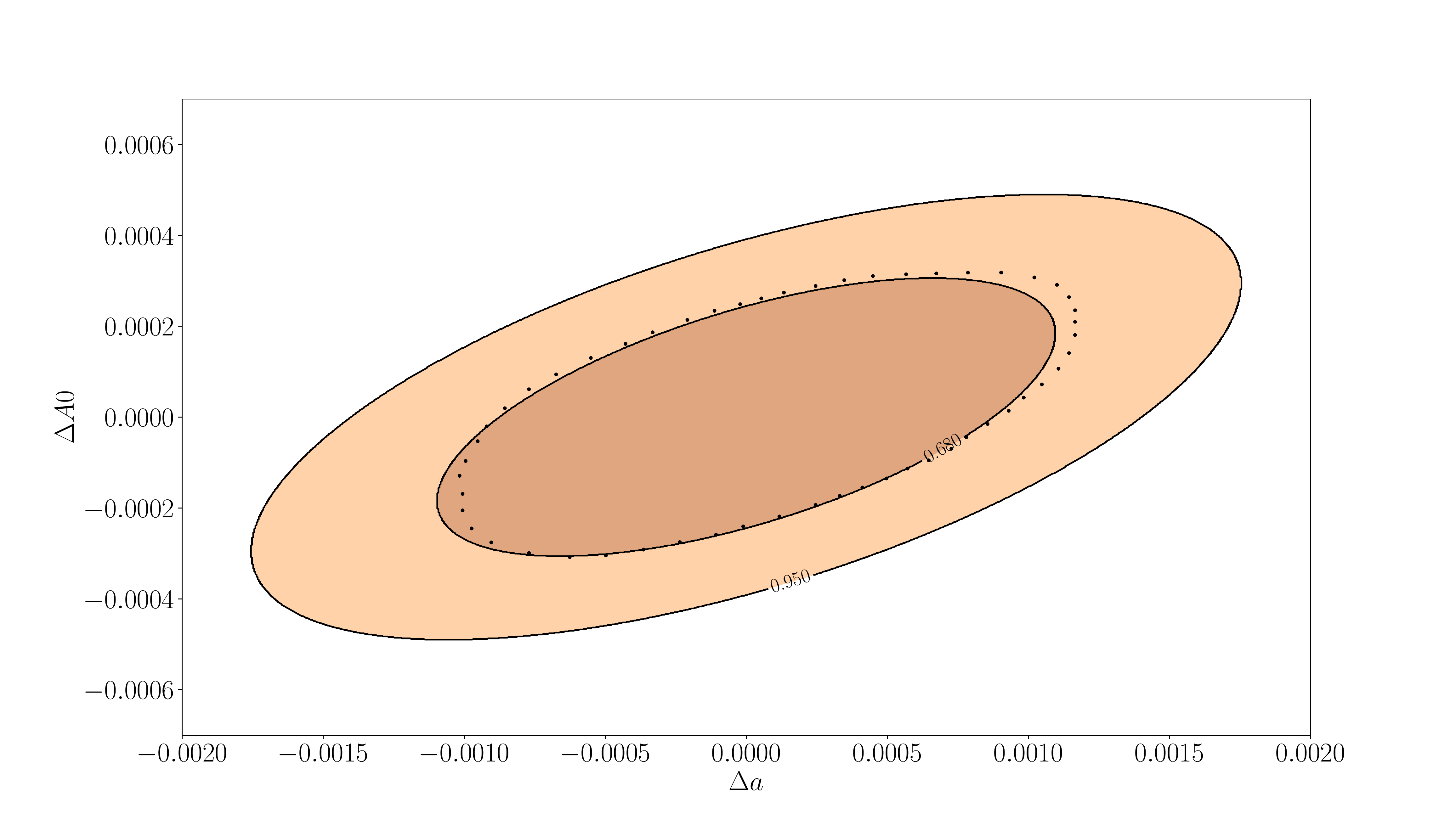}
\caption{The $68$ and $95\%$ confidence contours on the parameters $a$ and $A_0$, from the best-fit Gaussian to the MCMC chains provided by \cite{Despali_2015}. The black dots delineate the original $68\%$ contour from the MCMC chains.}
\label{aA0}
\end{figure}
%--------------------------------------------------------
Using the MCMC chains provided by \cite{Despali_2015}, we fit a 2-D Gaussian on the HMF parameter distribution, for each pair of parameters. We show an example in Figure~\ref{aA0} for the parameters $a$ and $A_0$. This enables us to reconstruct the full covariance matrix $C_\mathrm{HMF}$ for the HMF parameters, which we provide in Table~\ref{cov}.

\begin{table}
\caption{Covariance matrix, $C_{\mathrm{HMF}}$, reconstructed for parameters $(a,p,A_0)$, all dimensionless.}             % title of Table
\label{cov}      % is used to refer this table in the text
\centering                          % used for centering table
\begin{tabular}{c | c c c}        % centered columns (4 columns)
\hline\hline                 % inserts double horizontal lines
&$ a $& $p$ & $A_0$  \\    % table heading 
\hline                        % inserts single horizontal line
  $a$ & $5.56 \times 10^{-7}$ & $-1.01 \times 10^{-6}$ & $9.00 \times 10^{-8}$  \\      % inserting body of the table
  $p$ & $-1.01 \times 10^{-6}$ &  $3.19 \times 10^{-6}$  & $-1.93 \times 10^{-7}$ \\
  $A_0$ & $9.00 \times 10^{-8}$ & $-1.93 \times 10^{-7}$ & $4.37 \times 10^{-8}$ \\
\hline
\hline                                   %inserts single line
\end{tabular}
\end{table}
From the likelihood described in the previous section, one can compute the associated Fisher matrix for the clusters:
\begin{equation}
\label{fisher}
\quad\quad\quad\quad\quad\quad\quad\quad F_{\mathrm{clusters},{i,j}}=-\left\langle\frac{\partial^2 \ln \mathcal{L}}{\partial p_i \partial p_j}\right\rangle
\end{equation}
where $p_i\in \Theta\cup\Xi$ and the angle brackets indicate the average over the data ensemble corresponding to the fiducial parameter values.  The covariance matrix of the parameters $p_i$ is the inverse of the Fisher matrix.

We build the total Fisher information matrix as the sum of three components:
\begin{itemize}
\item The information obtained from cluster number counts, $F_\mathrm{clusters}$, given in Eq.~(\ref{fisher}),
\item The cosmological priors from \Planck\ cosmic microwave background measurements, $F_\mathrm{cmb}$~\citep{2018arXiv180706209P},
\item The HMF priors, $F_\mathrm{HMF}$.
\end{itemize}
The full Fisher matrix is then given by
\begin{equation}
\label{total_fish}
    F_\mathrm{tot}=F_\mathrm{clusters}+F_\mathrm{cmb}+F_\mathrm{HMF},
\end{equation}
where $F_\mathrm{cmb}$ (resp. $F_\mathrm{HMF}$) is non zero only in the $\Theta$ (resp. $\Xi$) quadrant. The non zero quadrant of $F_\mathrm{HMF}$ is the inverse of $C_\mathrm{HMF}$.

In this work, we examine the impact of $F_\mathrm{HMF}$ on cosmological parameter determination. In order to do so, we introduce the parameter $\alpha$, and we modify the priors on the HMF in Eq. (\ref{total_fish}) as follows: 
\begin{equation*}
    F_\mathrm{HMF}\leftarrow\frac{1}{\alpha}F_\mathrm{HMF}.
\end{equation*}
The covariance matrix being the inverse of the Fisher information, increasing $\alpha$ degrades the constraints on the HMF. On the other hand, $\alpha\rightarrow 0$ describes the situation where the HMF parameters are perfectly known. For $p\in\Xi$, this yields
\begin{equation*}
    \sigma_p^2\leftarrow\alpha\sigma_p^2,
\end{equation*}
where $\sigma_p$ is the error on one parameter of the HMF.
%-------------------------------------
\begin{figure}
\includegraphics[width=9cm,height=9cm]{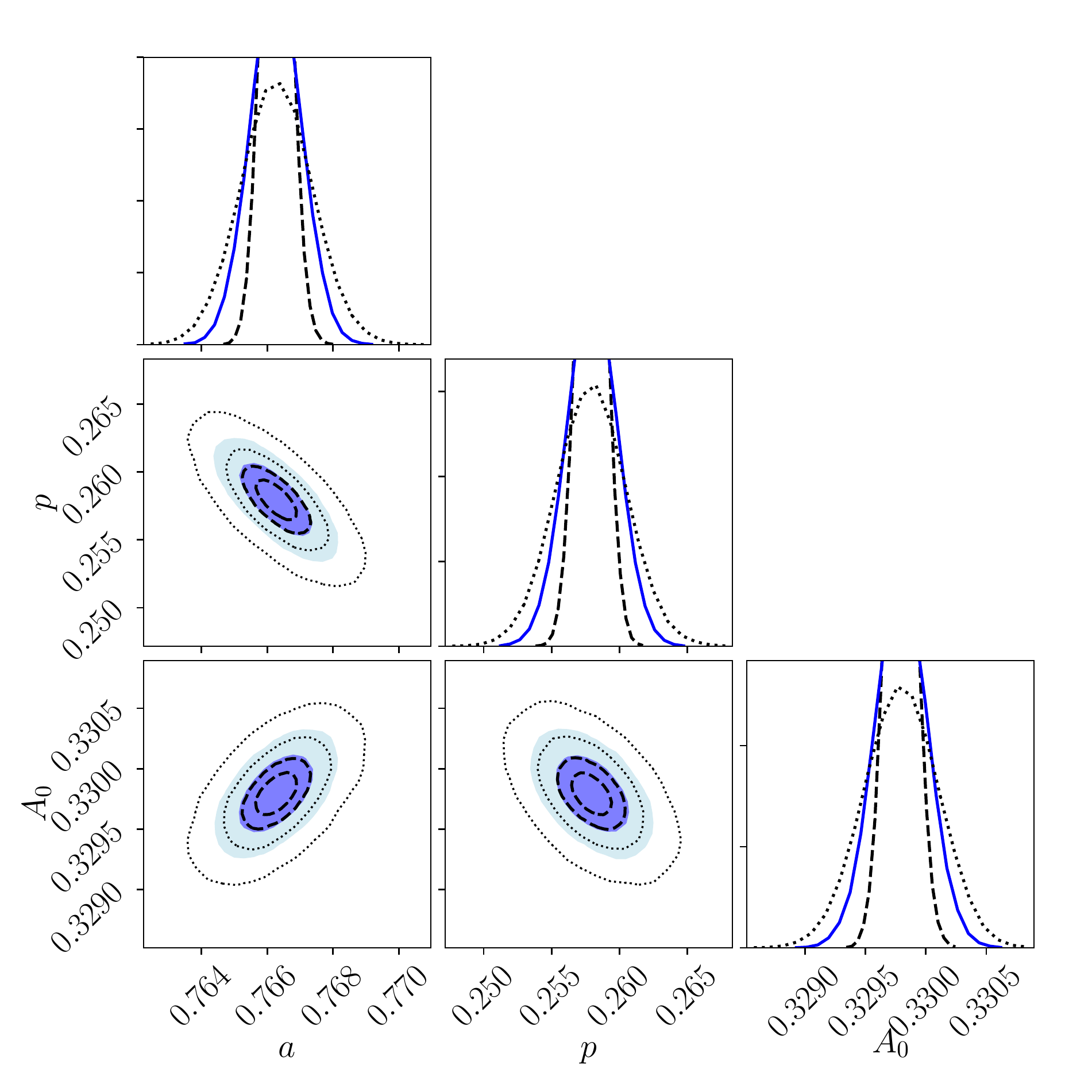}
\caption{The $68$ and $95\%$ percent contours on the parameters $a,p$ and $A_0$. The \textcolor{blue}{blue} ellipses are obtained when $\alpha=1$, i.e., the current precision provided by \cite{Despali_2015}. The black dashed curves are obtained for $\alpha=0.3$, and the dotted curves for $\alpha=2$.}
\label{diff_alpha}
\end{figure}
%-------------------------------------
Figure \ref{diff_alpha} illustrates how two values of $\alpha$ impact the confidence ellipses for the HMF parameters. 

Finally, we obtain
\begin{equation}
    \label{fisher_tot_alpha}
    F_\mathrm{tot}(\alpha)=F_\mathrm{clusters}(\alpha=1)+F_\mathrm{cmb}+\left(\frac{1}{\alpha}-1\right)F_\mathrm{HMF}.
\end{equation}
Here, the last term comes from the fact that the cosmological parameters are degenerate with the HMF for cluster cosmology. %, as illustrated by figure \ref{large_cat_mc}.
To get $F_\mathrm{clusters}$, we need to set a value for $\alpha$ (here $\alpha=1$) and remove it afterwards.
In the following sections, we change the value of $\alpha$ to measure its impact on  $\sigma_{p}$ for any $p\in\Theta$, the set of the cosmological parameters, and to identify the value necessary to avoid degrading the cosmological parameter constraints by more than $5\%$, and the total FoM by more than $10\%$.\\

\subsection{Cosmological models and mock catalogs}
\label{sec:studycases}

We generate catalogs for three different cosmological scenarios. Each model assumes a flat cosmology:
\begin{itemize}
    \item A base $\mathrm{\Lambda \mathrm{CDM}}$ cosmology (5 cosmological parameters\footnote{We do not consider the optical depth to reionization, $\tau$, as a parameter, since it has no impact on cluster counts, the normalization of the primordial power spectrum being set with $\sig$.}, $\Om$, $\sig$, $\Ho$, $\Ob$, $\ns$);
    \item A $w_0w_a$CDM cosmology, where we assume a dynamical dark energy model (7 parameters, the parameters from the base flat $\mathrm{\Lambda \mathrm{CDM}}$ and $\wo$, $\wa$ describing the dark energy equation-of-state \citep{2001IJMPD..10..213C}: $w(a)=\wo+\wa(1-a)$, where $a$ is the expansion scale factor normalized to unity at $z=0$);
    \item A $\nu$CDM cosmology, in which we consider the physics of massive neutrinos. In this case, the parameters are the same as the ones presented in the base $\mathrm{\Lambda \mathrm{CDM}}$, but we need our computation of the power spectrum, as described in section \ref{sec:nucdm}  (6 parameters, the parameters from the base flat $\mathrm{\Lambda \mathrm{CDM}}$ and $\smnu$).
\end{itemize}

The expected cluster counts associated with these models are computed with the python package HMF from \cite{murray2013hmfcalc}, which we have modified for the two latter cosmological scenarios in a way described below. The values adopted for the cosmological parameters for each scenario can be found in Table~\ref{fid_cosmology}.
\begin{table*}
\caption{Fiducial cosmological parameters adopted to generate the mock catalogs in the three cosmological scenarios. We adopt $\Tcmb=2.725 \, \mathrm{K}$, $\neff=3.04$ and $\smnu=0.06 \, \mathrm{eV}$ in the three cases. The parameters in black are the mean values of the chains provided by \cite{2018arXiv180706209P}. In \textcolor{red}{red}, the dark energy equation of state parameters, which are fixed for a $\Lambda$CDM cosmology.}
\label{fid_cosmology}      
\centering          
\begin{tabular}{c c c c c c c}     % 7 columns 
\hline\hline  
$\Om$ & $\sig$ & $\Ho ~ (\mathrm{km.s^{-1}.Mpc^{-1}})$ & $\Ob$ & $\ns$ & $\wo$ & $\wa$ \\ 
\hline     
                      % To combine 4 columns into a single one 
                  \multicolumn{7}{c}{$\Lambda$CDM (Planck2018) \& $\nu$CDM}\\
                                      
   0.3153 & 0.8111 & 67.36 & 0.04931 & 0.9649 & \textcolor{red}{-1} & \textcolor{red}{0} \\  
\hline  
                  \multicolumn{7}{c}{$w_0w_a$CDM (Planck2018)}\\
   0.2976 & 0.836 & 69.39 & 0.04645 & 0.9646 & -0.99 & -0.29 \\  
\hline                  
\end{tabular}
\end{table*}
They are the mean values of the chains provided by~\cite{2018arXiv180706209P}: PklimHM-TT-TE-EE-lowL-lowE-lensing chains in the $\Lambda$CDM and $\nu$CDM cases and PklimHMTT-TE-EE-lowL-lowE-BA0-Riess18-Panthe for $w_0w_a$.  For the case of the $\nu$CDM scenario, we adopt the lower limit provided by \cite{Abe_2008}, $\smnu = 0.06 \, {\rm eV}$, because the best fit value provided by \Planck\ is lower.

The current lack of a precise understanding of the selection function of future optical surveys leads us to consider simple mass and redshift thresholds for the completeness of the catalog:
\begin{equation}
\label{self_}
\chi(M,z)=\left\{
    \begin{array}{ll}
        1 & \mbox{if } M \in [M_{\mathrm{thres}},+\infty] ~ \mbox{and}~ z\in[0,2] \\
        0 & \mbox{otherwise}
    \end{array}
\right.
\end{equation}
The total expected number of objects in the catalog is then
\begin{equation}
\label{tot}
N_{tot}=\int_{M_{\rm thres}}^{\infty}\int_{z=0}^{2} \partial_{\small{{z,\ln{M}}}}N(M,z) \mathrm{d}\ln M \mathrm{d}z.
\end{equation}
 We set the mass threshold so that $N_{tot}$ is close to the number of clusters expected with a survey like \Euclid, i.e., $\sim$$60,000$ clusters over $15,000 \, \mathrm{deg}^2$~\citep{2016MNRAS.459.1764S}. For the DMF, this mass threshold is $\log (M_{\rm thres}/M_\odot )=14.18$. We then reduce/increase the area of the survey to obtain $N_{tot}$ of $\sim$$500$/$\sim$$1.5 \times 10^5$ clusters (\Planck-like and Future projects).
 
 We generate five cluster catalogues. The first three ($\mathrm{C}_{\mathrm{Euclid}}$, $\mathrm{C}_{\mathrm{Planck}}$, $\mathrm{C}_{\mathrm{Future}}$) fix the sum of the neutrino masses to its minimal value of $\sum m_\nu=0.06$ and are used for analyses in the $\Lambda$CDM and $w_0w_a$CDM scenarios. Since constraining the masses of the neutrinos requires a special treatment for the power spectrum (see Sect.~\ref{sec:nucdm} for details), we generate two additional catalogs for the $\nu$CDM scenario corresponding to \Euclid-like and Future surveys. Since we are interested in forecasting the precision on the measurement of the neutrino mass scale, we do not generate a \Planck-like catalog for the $\nu$CDM scenario. 
 The mocks are obtained from a Poisson draw of the DMF. The characteristics of each catalog are given in~Table~\ref{cat_chars}.\\
 \begin{table}
\caption{Characteristics of the mock catalogs. The first column is the survey area. Note that varying neutrino mass in the analysis changes the power spectrum. Thus, we generate a special set of two catalogs when letting the neutrino mass free. The two following columns are the number of clusters present in the catalog for a $\Lambda$CDM and a $w_0w_a$CDM cosmology, and the last one is the same for $\nu$CDM (The cosmological parameters correspond to Table \ref{fid_cosmology}). For all the catalogs, $[z_{min},z_{max}]=[0,2]$ and $M_{\mathrm{thres}}(\log(M/M_\odot))=14.18$.}             % title of Table
\label{cat_chars}      % is used to refer this table in the text
\centering                          % used for centering table
\begin{tabular}{c c c c c}        % centered columns (4 columns)
\hline\hline                 % inserts double horizontal lines
&$\Omega$($\mathrm{deg}^2$) & $N_{\mathrm{clusters}}$  & $N_{\mathrm{clusters}}$ &
$N_{\mathrm{clusters}}$\\    % table heading 
&     & $\Lambda$CDM & $w_0w_a$CDM & $\nu$CDM \\
\hline                        % inserts single horizontal line
  $\mathrm{C}_{\mathrm{Future}}$ & 34,503 & 153,891 & 191,408 & 151,427\\  % inserting body of the table
  $\mathrm{C}_{\mathrm{Euclid}}$ & 15,000 & 66,252 & 82,694 & 65,994 \\
  $\mathrm{C}_{\mathrm{Planck}}$ & 114   & 513 &617& - \\
\hline                                   %inserts single line
\end{tabular}
\end{table}

\section{Results}
\label{sec:results}

We provide here the results for the three cosmological scenarios ($\Lambda$CDM, $w_0w_a$CDM, $\nu$CDM) and the three catalogs (\Planck-like, \Euclid-like and Future) whose characteristics are given in the previous section. For each scenario and each catalog, we run the same analysis, and we give the precision required on the halo mass function. 

Parameters which are poorly constrainted with cluster counts such as $\Ob$, $\ns$, $\Ho$ make the computation of the coefficients of the cluster Fisher matrix $F_\mathrm{clusters}$ difficult. Indeed, since the log-likelihood function is almost constant, we are more subject to numerical instabilities when we compute the second order derivatives.
However, our analysis requires that we are able to capture sufficiently the correlations between the different parameters.
We thus decided to run first an MCMC analysis and fit a Gaussian on the contours as we did for the HMF parameters in Sect.~\ref{sec:fisher}. The Fisher matrix is then deduced as the inverse of the covariance matrix. The MCMC analysis is performed with the emcee python package from \cite{Foreman_Mackey_2013}.
Large flat priors are applied for the cosmological parameters of interest. They are summarized in Table~\ref{priorslcdm}.
\begin{table}
\caption{Priors applied on the flat $\Lambda$CDM cosmology}             % title of Table
\label{priorslcdm}      % is used to refer this table in the text
\centering                          % used for centering table
\begin{tabular}{c c c }        % centered columns (4 columns)
\hline\hline                 % inserts double horizontal lines
Parameter & Prior & Range \\    % table heading 
\hline                        % inserts single horizontal line
   $\Omega_m$ & Flat & $[0.1,0.7]$  \\      % inserting body of the table
   $\sigma_8$ & Flat  & $[0.3,0.99]$    \\
   $H_0$ & Flat  & $[40,250]$      \\
   $\Omega_b$ & Flat  & $[0.03,0.06]$  \\
   $n_s$ & Flat  & $[0.5,1.7]$     \\ 
   $a$ & Gaussian & from $\alpha \times C_\mathrm{HMF}$    \\ 
   $p$ & Gaussian &  from $\alpha \times  C_\mathrm{HMF}$  \\ 
   $A_0$ & Gaussian & from $\alpha \times C_\mathrm{HMF}$  \\ 
\hline                                   %inserts single line
\end{tabular}
\end{table}
Convergence is checked with the integrated autocorrelation time, as suggested by \cite{2010CAMCS...5...65G}.

\subsection{$\Lambda$CDM cosmology}

We present the results for the clusters-only case, $F_\mathrm{clusters}(\alpha=1)+\left(\frac{1}{\alpha}-1\right)F_\mathrm{HMF}$, in Sect.~\ref{sec:lcdm_clusters_only} and the \mbox{clusters + \Planck\ CMB}-case, $F_\mathrm{clusters}(\alpha=1)+F_\mathrm{cmb}+\left(\frac{1}{\alpha}-1\right)F_\mathrm{HMF}$, in Sect.~\ref{sec:lcdm_clusters_planck}.

\subsubsection{Cosmological constrains from clusters only}
\label{sec:lcdm_clusters_only}

\begin{figure*}
\includegraphics[width=\textwidth,height=23cm]{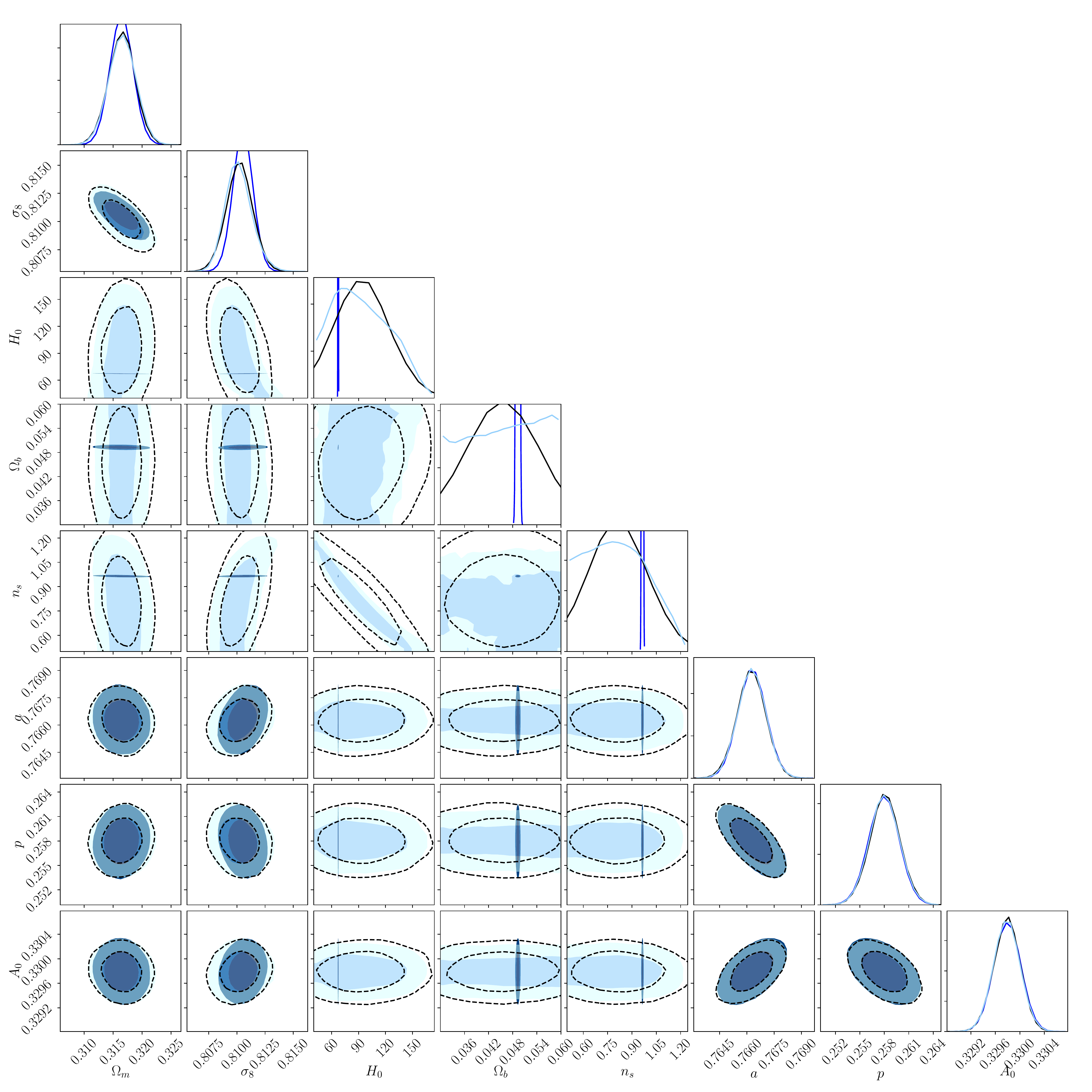}
\centering
\caption{\small{In light blue, MCMC $68\%~\textrm{and}~95\%$ contours for the cosmological parameters and the halo mass function parameters for a \Euclid-like catalog. We fixed $\alpha=1$. The dark blue contours are obtained when adding \Planck\ priors. The dashed lines are the Gaussian fit from which the covariance matrix and subsequently the Fisher matrix $F_\mathrm{clusters}(\alpha=1)$ is computed.}}
\label{euclid_gauss}
\end{figure*}

Since galaxy cluster number counts are particularly sensitive to $\Om$, and $\sig$, we naturally focus on these parameters for our analysis. In Figure \ref{euclid_gauss}, we show the cosmological constraints for the \Euclid-like catalog with $\alpha=1$, as well as our Gaussian approximation. As expected, $\Ob$ is poorly constrained as the clusters primarily probe the total matter content of the universe.

We now vary the parameter $\alpha$ encapsulating the uncertainty on the HMF parameters to see if we reach the required precision on the cosmological parameters. Recall that we require $5\%$ precision on individual cosmological parameters (here, $\Om$ and  $\sig$), and $10\%$ precision on the total FoM (here, in the ($\Om$,$\sig$)-plane).
\begin{figure}
\includegraphics[width=9cm]{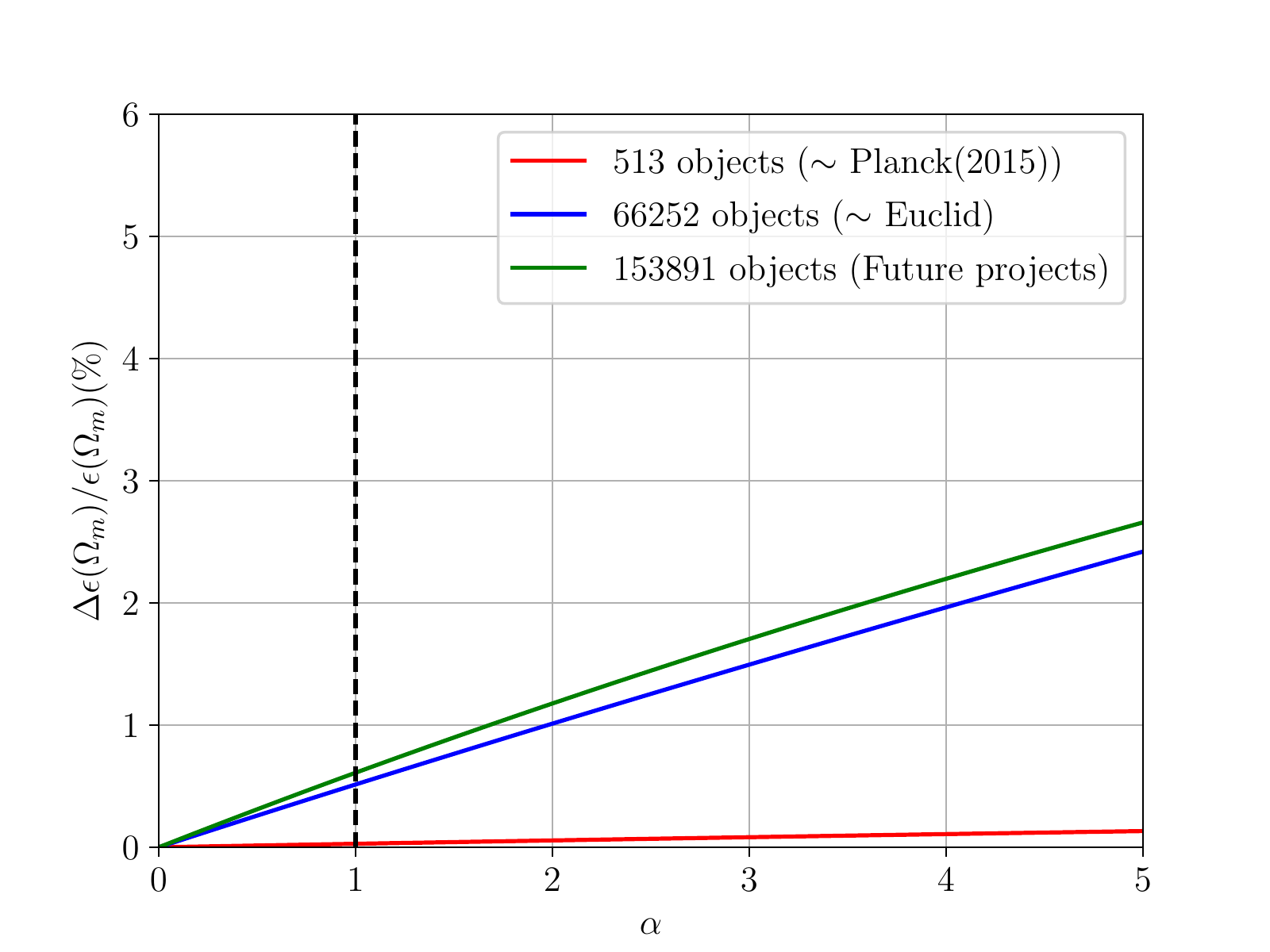}
\caption{Relative variation of the error on the matter density parameter $\epsilon(\Om)$ for the three different sizes of catalogs. For all catalogs, the value of $\alpha$ required to avoid degrading the constraint by more than 5\% is much greater than 1. The current knowledge of the mass function ($\alpha=1$) is sufficient. This is the result for clusters only; the results for clusters plus \Planck\ 2018 priors are summarized in Table~\ref{reslcdm}.}.
\label{omlcdm}
\end{figure}

Figure \ref{omlcdm} shows that the precision on the parameter $\Om$ is not strongly affected by the variation of the parameters of the mass function. This is expected since Figure \ref{euclid_gauss} shows that this parameters is poorly correlated with the HMF parameters. 

\begin{figure}
\includegraphics[width=9.5cm]{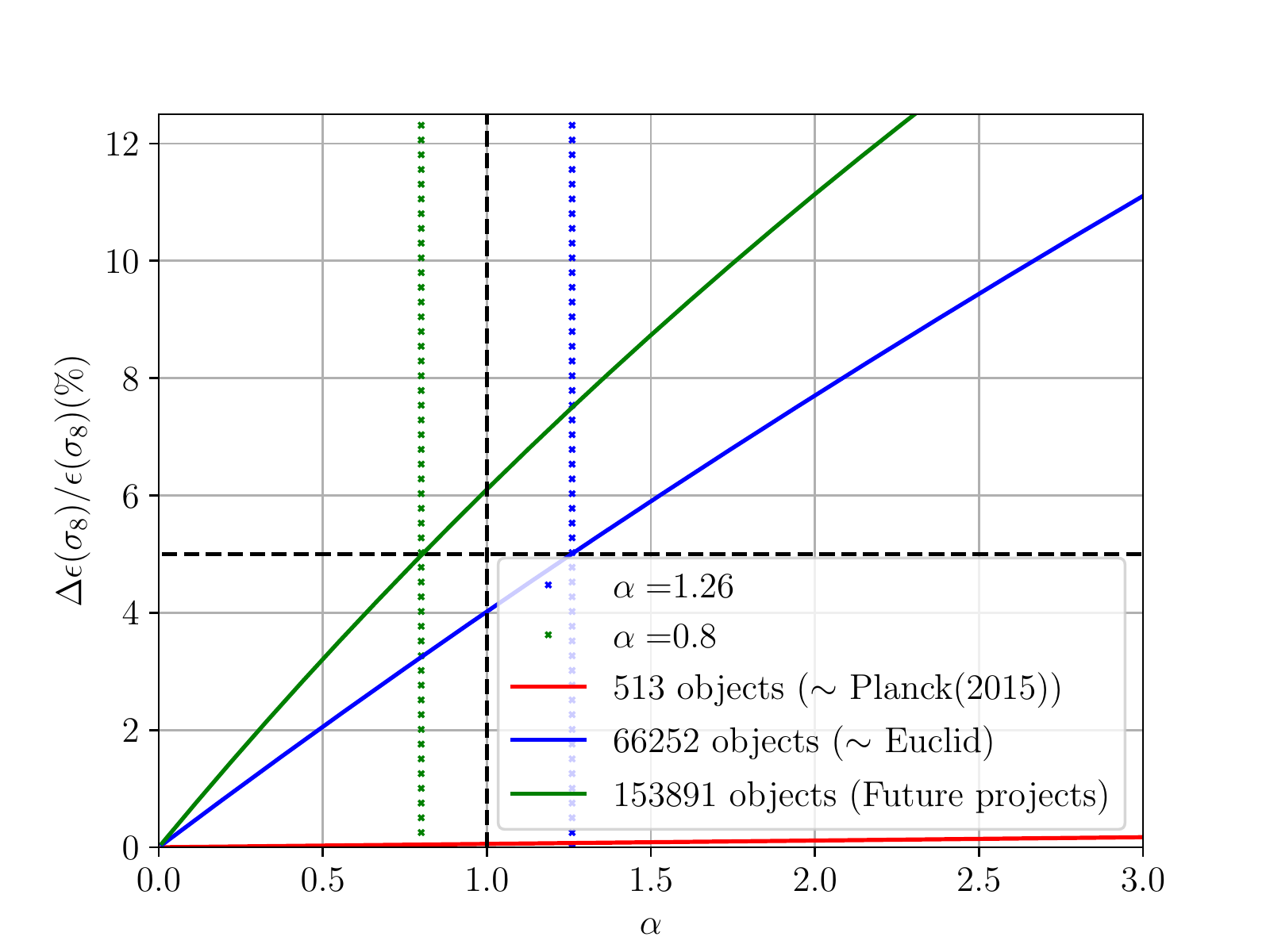}
\caption{Relative variation of the error on $\sig$ for the three catalogs. The level of precision on the HMF reaches the limit of  current knowledge ($\alpha\rightarrow 1$) for a \Euclid-like catalog (\textcolor{blue}{Blue}). For the largest catalog (\textcolor{green}{Green}), the constraints on the HMF need to be $20\%$ better.  This is the result for clusters only; the results for clusters plus \Planck\ 2018 priors are summarized in Table~\ref{reslcdm}.}.
\label{siglcdm}
\end{figure}

On the other hand, $\sig$ is correlated with $a, p$ and $A_0$. Figure~\ref{siglcdm} shows that, for this parameter and a Future cluster catalog, we need better constraints (20\% improvement) on the halo mass function parameters than those currently available.

\begin{figure}

\includegraphics[width=9.5cm]{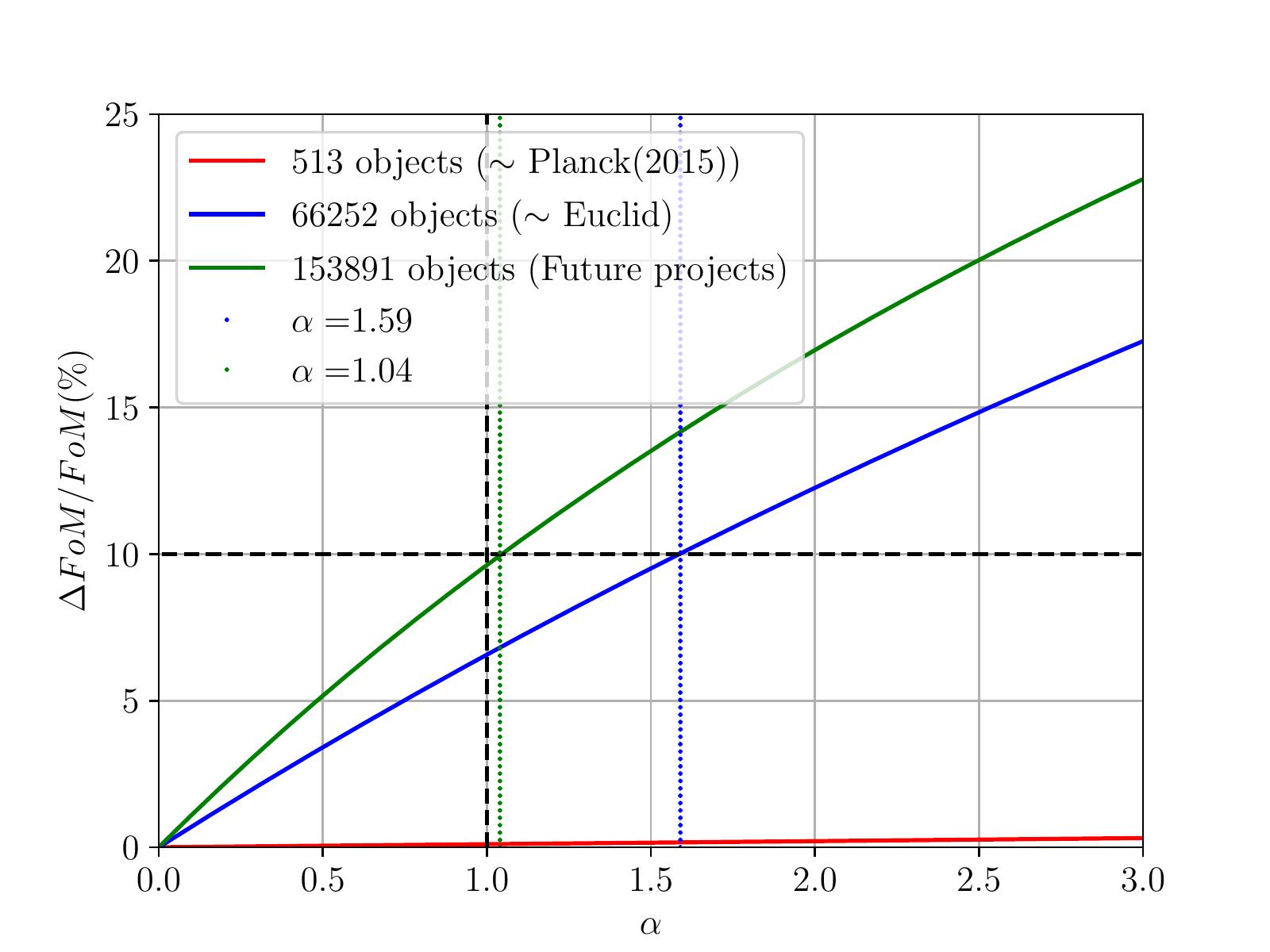}
\caption{Relative variation of the FoM (inverse of the area within the error ellipse in the two-dimensional parameter space) in the $\Om-\sig$ plane. For the two large catalogs, $\alpha$ is close to the limit value 1.  This is the result for clusters only; the results for clusters plus \Planck\ 2018 priors are summarized in Table~\ref{reslcdm}.}.
\label{fomlcdm}
\end{figure}

The last issue concerns the FoM. Figure~\ref{fomlcdm} demonstrates that, for clusters only, for a catalog of the size of \Euclid\ and for a Future catalog, the current precision is sufficient to match our objective of $10\%$. 

\subsubsection{Cosmological constrains from clusters with Planck 2018 priors}
\label{sec:lcdm_clusters_planck}

Since clusters only poorly constrain $\Ob, \Ho$ and $\ns$, we will combine them with external constraints. The most natural complementary probe is the \Planck\ primary CMB~\citep{2018arXiv180706209P}. 
Consequently, we include priors from \Planck\ as $F_\mathrm{cmb}$ (see Sect.~\ref{sec:fisher}) in our total Fisher matrix, and we run our analysis a second time. The priors are the same as in the previous section (see Table~\ref{priorslcdm}). For clarity, we do not add figures similar to those presented in the previous section, but instead summarize our results in Table~\ref{reslcdm}. Corner plots for cosmological and HMF parameters are given in Appendix~\ref{app:lcdm}.

For $\Om$ the impact is more important because the constraints are tighter, but the current level of precision on the HMF is nevertheless good enough to avoid degrading the constraints beyond our threshold of 5\%.
However, this time $\sigma_8$ and the FoM are strongly affected, and improvements in HFM precision are needed to avoid degrading these constraints. 
All the results are provided in Table~\ref{reslcdm}. We highlighted in red when the precision on the HMF needs to be improved. The precision on the HMF parameters needs to be improved for \Euclid-like and Future catalogs (up to $\sim70\%$ for the FoM for Future catalogs) if one wants to fully exploit their cosmological potential.

\begin{table}
\caption{Maximum value allowed for the $\alpha$ parameter in Flat $\Lambda$CDM. Critical cases requiring improved precision on HMF parameters correspond to a value lower than 1 and are highlighted in \textcolor{red}{red}.}             % title of Table
\label{reslcdm}      % is used to refer this table in the text
\centering                          % used for centering table
\begin{tabular}{c c c }        % centered columns (4 columns)
\hline\hline                 % inserts double horizontal lines
       &     Clusters only      &    Clusters+Planck2018 \\    % table heading 
\hline                      % inserts single horizontal line
   \Planck-like &    &    \\ 
   (513 objects) &     &     \\
   $\Om$ & $\gg1$ & $\gg1 $  \\ 
   $\sig$ & $\gg1$ & $\gg1$ \\
    FoM & $\gg1$ & $\gg1$ \\
   \hline     % inserting body of the table
   \Euclid-like &  &   \\ 
   (66,252 objects) &  &  \\
    $\Om$ & $\gg1$ & $\gg1$  \\ 
   $\sig$ & $ 1.2 $ & $\textcolor{red}{0.8}$ \\
    FoM & $1.6$ & $\textcolor{red}{0.67}$ \\
   \hline     % inserting body of the table
  Future &  &   \\ 
   (153,891 objects) &  &  \\
    $\Om$ & $\gg1$ & $ 3.6 $  \\ 
   $\sig$ & \textcolor{red}{0.8} & $\textcolor{red}{0.32}$ \\
    FoM & $1$ & $\textcolor{red}{0.31}$ \\
\hline                                   %inserts single line
\end{tabular}
\end{table}

\subsection{$w_0w_a$CDM}
The most common way of modeling the accelerated expansion of the universe is to consider a fluid with negative pressure, which is characterized by its equation-of-state $p=w\rho$. A cosmological constant corresponds to $w=-1$ and $\rho=\mathrm{cst}$. Even though the  cosmological constant is consistent with current observations, many different dark energy models predict a dynamical equation-of-state. One of the most popular parametrizations was proposed by  \cite{2001IJMPD..10..213C} and \cite{2003PhRvL..90i1301L},
$$w(a)=w_0+w_a(1-a),$$
which, as a two parameter expansion, adds a layer of complexity. 

In this model, the HMF parameters have a different effect. First, leaving $w_0$ and $w_a$ free increases the size of the confidence contours. In addition, the shape of the $68~\mathrm{and}~95\%$ regions changes (see Fig.~\ref{euclid_wcdm} and~\ref{futur_wcdm}). Corner plots for cosmological and HMF parameters are provided in Appendix~\ref{app:wowacdm}.

For these reasons, the impact of the HMF parameters is smaller than in the $\Lambda$CDM case. For all three catalogs, we never reached a loss of relative precision comparable to the $\Lambda$CDM case, not only for $\Om$ and $\sig$, but also for the dark energy equation-of-state parameters $w_0$ and $w_a$. This should not obscure the fact that these results are highly dependant on the  HMF uncertainties.  Moreover, future  experiments (CMB, BAO, etc) will constrain cosmology better than the \Planck ~ priors  used here. These two aspects are discussed further in Sect.~\ref{discussion}. 

\subsection{$\nu$CDM}
\label{sec:nucdm}
For our final study we  vary the the sum of the neutrino masses. The presence of massive neutrinos induces a scale dependence of the growth factor, and slightly changes the model. Following  \cite{Costanzi_2013}, we model the power spectrum as
\begin{equation}
    P_{\mathrm{cdm}}(k,z)=P_{\mathrm{m}}(k,z)\left(
    \frac{\Om T_\mathrm{cdm}(k,z)+\Ob T_\mathrm{b}(k,z)}{T_\mathrm{m}(k,z)(\Omega_\mathrm{cdm}+\Ob)}\right)^2,
\end{equation}
and we also change the collapse threshold \citep{Hagstotz_2019}
\begin{equation}
    \delta_c^\nu=\frac{\sigma(z)}{\sigma_{\mathrm{cdm}}(z)}\delta_c.
\end{equation}

With this formalism, we forecast  cosmological constraints from the two mock catalogs generated to study varying neutrino mass, as described in Sect.~\ref{sec:studycases}. As we have already noted, the HMF uncertainty is not significant compared to the statistical uncertainties for a \Planck-like catalog.

For the primary CMB prior $F_{\rm cmb}$, we take the \Planck\ chains $\mathrm{PklimHM}+\mathrm{TT}+\mathrm{TE}+\mathrm{EE}+\mathrm{lowL}+\mathrm{lowE}+\mathrm{lensing}+\mathrm{BAO}$ with neutrino mass as a free parameter.  Since the best fit value for the total neutrino mass provided by these chains is lower than the minimal value required by neutrino oscillation experiments~\citep{Abe_2008}, we center the \Planck\ Fisher contours on the fiducial value of $0.06 \, {\rm eV}$, the minimal mass value.

\begin{figure}%[H]
\includegraphics[width=9.5cm]{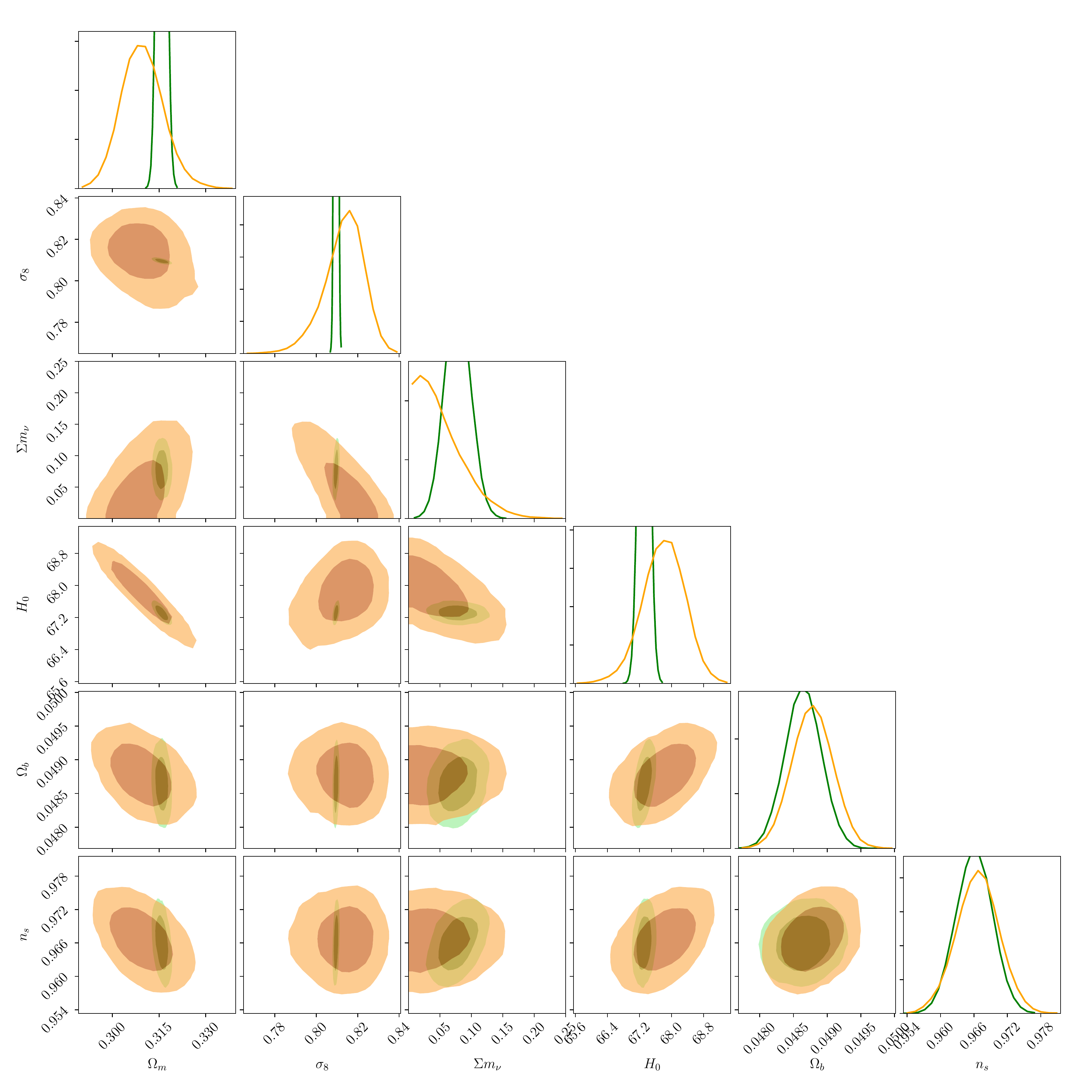}
\caption{Cosmological parameters obtained for a catalog of 151,427 objects (Future). In \textcolor{orange}{orange}, the \Planck\ (2018) contours, and in \textcolor{green}{green}, the clusters combined with \Planck\ priors. We expect, for the largest catalog, a $2\sigma$ measurement of the neutrino mass scale. The full corner plot with all parameters is given in Appendix~\ref{app:nucdm}.}
\label{cluster_futur_planck_nu}
\end{figure}
 Figure \ref{cluster_futur_planck_nu} shows the results for the largest catalog. We expect to achieve a $2\sigma$ measurement of the neutrino mass scale. Full corner plots for both cosmological and HMF parameters are provided in Appendix~\ref{app:nucdm}. %Fig.~\ref{future_nucdm}
 We see that clusters significantly improve the constraints on neutrino mass when added to the current \Planck\ priors, in particular by constraining the lower mass range. And if neutrinos are slightly more massive than the minimal value, the confidence limits would improve: The green contour in the $\Sigma m_\nu-\sig$ plane would move to  higher values, but since the CMB would still constrain the the upper value of the neutrino mass, we would obtain notably tighter constrains. In this sense, the plot is pessimistic.  Moreover, the result shown here marginalizes over the \Planck\ uncertainty on the optical depth to reionization, which can be expected to improve in the coming years and hence yield even tighter constraints on neutrino mass.\\
\begin{figure}
\includegraphics[width=9cm]{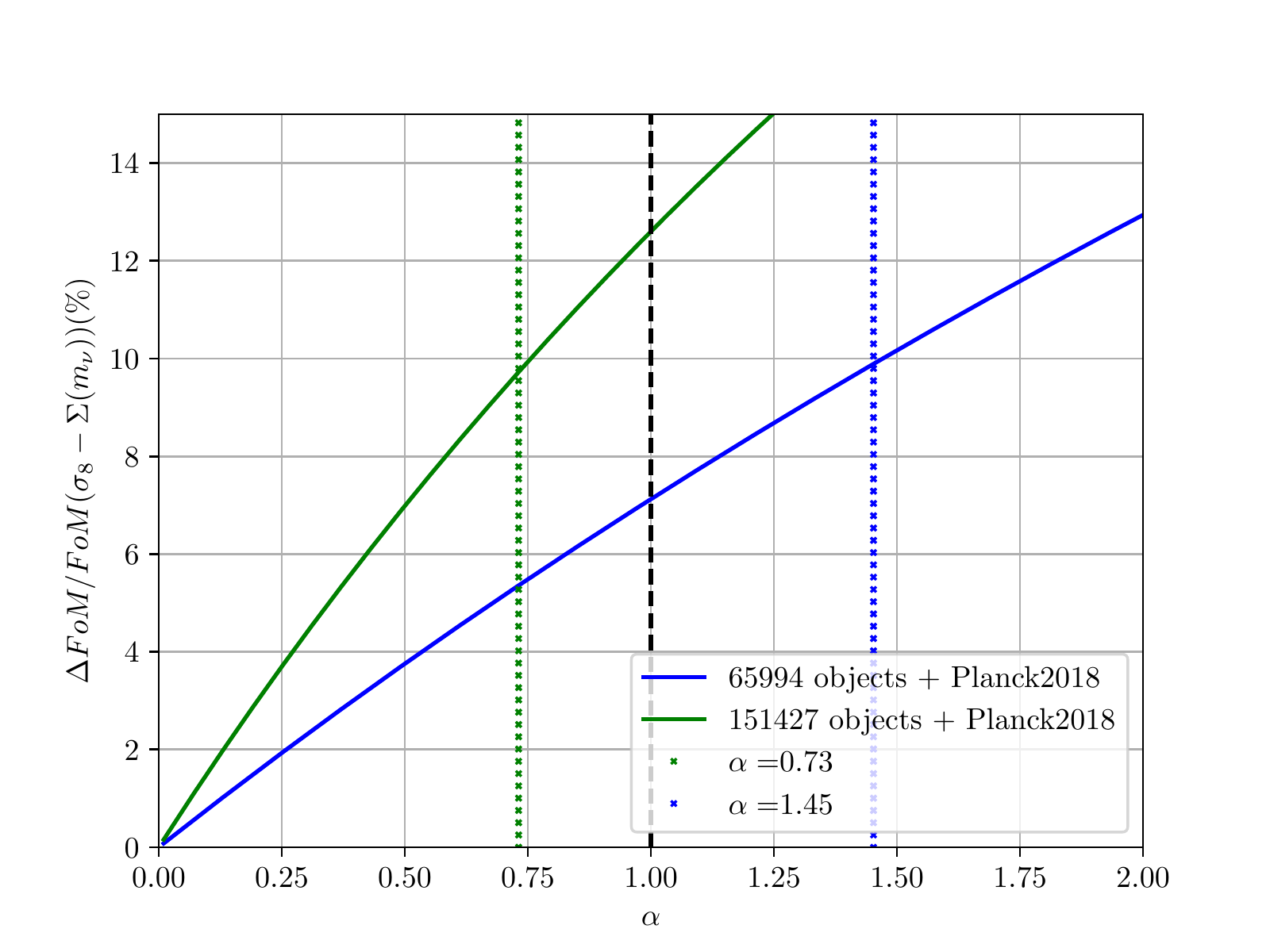}
\caption{Evolution of the uncertainty in the $\Om-\smnu$ plane. In \textcolor{blue}{blue}, the \Euclid-like case, and in \textcolor{green}{green} a larger Future catalog.}
\label{error_evol_sig_mnu}
\end{figure}

We perform the same analysis as the previous section, focusing only on the Clusters+Planck2018 case. Results are summarized in Table~\ref{res_nucdm}. We observe the same effect as for the $w_0w_a$CDM scenario: since we allow the chains to explore a larger area of parameter space, it appears that the impact of our knowledge of the HMF is less important than in the case of $\Lambda$CDM.  For a \Euclid-like catalog, the current precision on the HMF parameter is sufficient, but not for the case of a Future catalog, which will require improvements by $\sim40\%$ in order to not degrade the constraint on $\sig$ by more than $5\%$.

We must emphasize, however, that the best fit values of the HMF parameters $(a,p,A_0)$ (and associated uncertainties) are provided for a $\Lambda$CDM scenario.  We have  assumed these to be valid in the $\nu$CDM case, although we could reasonably expect the errors to be larger for a $\nu$CDM cosmology. We will come back to this point in the discussion.
\begin{table}
\caption{The maximum value allowed for  $\alpha$ in a $\nu$CDM cosmology. Critical cases requiring improved precision on HMF parameters correspond to a value lower than 1, and are highlighted in \textcolor{red}{red}.}             % title of Table
\label{res_nucdm}      % is used to refer this table in the text
\centering                          % used for centering table
\begin{tabular}{c c }        % centered columns (4 columns)
\hline\hline                 % inserts double horizontal lines
       &  Clusters+Planck2018   \\    % table heading 
\hline                      % inserts single horizontal line
   \Euclid-like &     \\ 
   (65,994 objects)   &  \\
    $\Om$ &  $ 7.7 $  \\ 
   $\sig$ &  1.03 \\
   $\smnu$ &  $\gg1$ \\
    FoM($\Om-\sig$) &  1.57 \\
    FoM($\sig-\smnu$) &  1.45 \\
   \hline     % inserting body of the table
  Future &    \\ 
   (151,427 objects)  &  \\
    $\Om$ & $ 3.86 $  \\ 
   $\sig$ &  $\textcolor{red}{0.61}$ \\
   $\smnu$ &  $\gg1$ \\
    FoM($\Om-\sig$) & $\textcolor{red}{0.85}$ \\
    FoM($\sig-\smnu$) &  $\textcolor{red}{0.73}$ \\
\hline                                   %inserts single line
\end{tabular}
\end{table}

\section{Summary of Results}
\label{sec:summary}
 
Our study focuses on cluster counts as a cosmological probe.  The wider and deeper surveys of the future will demand every increasing attention to systematics. 
As the primary theoretical link between cosmological quantities and the observed counts, the HMF is the mainstay of every past and future analysis. The most important aspects of the HMF in this context are:
\begin{itemize}
    \item Universality, or the concept that the HMF functional parameters are not explicitly dependant on redshift nor cosmology.
    \item  Accuracy, the ability to recover unbiased cosmological parameters.
    \item Precision in the fit of the HMF functional parameters when fit on numerical simulations. 
\end{itemize}

State-of-the-art simulations are improving our understanding of the first two aspects \citep{Despali_2015}.  In this paper, we focus on the last point, estimating the required precision on a given set of parameters in order to avoid degrading  cosmological constraints. We discuss the first two aspects in Section~\ref{discussion}.

For our purposes, we set a threshold of $5\%$ for allowed degradation in the precision for any single cosmological parameter, and $10\%$ for the FoM.
We find that, for the $\Lambda$CDM scenario, and for a \Euclid-like catalog when \Planck\ priors are applied, one of the principal parameter of interest, $\sig$, as well as the FoM are violate this threshold with the present precision: the precision on the HMF parameters must be improved by  $\sim30\%$ (Table \ref{reslcdm}). For a Future catalog, the situation is more critical: an improvement of $\sim70\%$ is required to avoid degrading the FoM on $(\Om,\sig)$ by more than 10\%.

For $w_0w_a$CDM, any degradation in the
dark energy equation-of-state parameters remains under a percent. 
This is simply because allowing these parameters to vary increases the size of the confidence ellipses, thereby reducing the impact of the HMF parameters. A look at the corner plots in Appendix~\ref{app:wowacdm} shows that there is in fact little correlation between ($w_0,w_a$) and ($a,p,A_0$) in this case.

Lastly, we examined constraints on the sum of neutrino masses. Once again for a future catalog, the current precision on HMF parameters does not meet our threshold requirement, and one needs to improve the constraints on the HMF parameters by $\sim30\%$ (Table \ref{res_nucdm}). For a \Euclid-like catalog, the current precision appears to be sufficient.

\section{Discussion \& Conclusion}
\label{discussion}

The HMF is the foundation of cosmology with cluster counts.  It promises to provide a robust, simple, analytical form for the abundance of dark matter halos across cosmological models as a function of linear-theory quantities only.  Universality is the key desired property: a functional form whose parameters do not explicitly depend on the cosmological scenario.  Cosmological dependence only enters through the linear-theory power spectrum and growth factor.  In this way, a universal HMF enables exploration of cosmological parameter space without the need for expensive numerical simulations at every step.
In most analyses to date, the HMF parameters are fitted on a set of numerical simulations, and then fixed in the cosmological analysis, for example \cite{2016A&A...594A..24P}.

Cluster catalogs will vastly increase in size with planned experiments, such as eROSITA (now operating), Simons Observatory, CMB-S4, \Euclid, \RomanT\ and \Rubin. As statistical power increases, we must quantify both the {\em accuracy} and the {\em precision} of the HMF fitting form, and evaluate their impact on cosmological constraints.  The former tells us to what extent the best-fit HMF functional form reproduces the halo abundance in simulations; the latter refers to the precision of the HMF parameters around their best-fit values.  This has not been an issue at the times when cluster catalogs were containing only a few hundreds objects. With \Euclid, we expect $10^5$ clusters, so these errors can no longer be ignored \citep{10.1093/mnras/stz1949}.

Our study provides a framework to estimate the impact of the precision of HMF parameters on cosmological parameter estimation. 
We do this by freeing the parameters of the HMF, applying priors from the fit and quantifying the impact on the cosmological inference.  We found that improvements of up to 70\% in the precision of HMF parameters is required to prevent degrading cosmological constraints by more than our threshold of 5\% on a single parameter and 10\% on the FoM.  

Our results are reliable only for the mass function from~\cite{Despali_2015}. The results could change for other HMFs depending on the correlations between HMF parameters and cosmological quantities. Thus, our main result is that a simlar discussion should be made when presenting error bars extracted from simulations. It will not be possible in future analyses to simply fix the best-fit parameters of the HMF when running a cosmological analysis. One will have to adjust at the same time cosmology, and halo mass function parameters, with priors from simulation for these latter.

No perfect functional form exists for the HMF, and in fact Fig.~\ref{comp_tinker_desp} shows that the error bar can not provide an overlap between two of the most widely used HMFs.  This is the issue of accuracy.  It is therefore essential to also verify that the adopted HMF provides unbiased cosmological constraints (see \cite{2020A&A...643A..20S}).  Differences in HMFs can arise from numerous sources, including the effect of baryons \citep{2021MNRAS.500.2316C}, which will be a major issue in the upcoming years.

A related issue is whether different cosmological scenarios provide different uncertainties on HMF parameters (apart from the best-fit values).  Indeed, the HMF formalism is based on the idea of universality, i.e., the claim that the HMF parameterization does not depend on the cosmology. \cite{Despali_2015} demonstrated that this is nearly the case over a broad range of cosmological scenarios. But this only means that the best-fit is the same for the HMF parameters in different scenarios. Nothing is said about the uncertainties.
There is no reason why the errors should not be affected when the cosmology changes, and this should be taken into account.

To summarize, we have to be sure that, in a given mass and redshift interval, the error bars provided encapsulate all the uncertainties related to simulation and halo finder (see \cite{2011MNRAS.415.2293K} for a review).

In the near future, we will have discovered almost all dark matter halos hosting massive clusters of galaxies. But as statistics increases, effort must be made to not only provide a good description of what is seen in the simulations, but also to provide a framework capable of yielding unbiased cosmological constraints. We have focused on this latter point, in the special case of halo mass function parameters. Our conclusion is that it is important to construct a likelihood where 
\begin{itemize}
\item Cosmology ($\Om,\sig,\wo,\wa,\smnu$,...),
\item Selection function with priors from detection algorithm \citep[see e.g.][]{2019A&A...627A..23E},
\item Halo mass function with priors from simulation,
\item Observable/mass relation with priors from simulation \citep[see e.g.][]{2019MNRAS.484.1598B},
\item Large scale bias of dark matter halos $b(M,z)$ with priors from simulation \citep[see e.g.][]{2010ApJ...724..878T}
\end{itemize}
are adjusted simultaneously.

\begin{acknowledgements}
A portion of the research described in this paper was carried out at the Jet Propulsion Laboratory, California Institute of Technology, under a contract with the National Aeronautics and Space Administration.
\end{acknowledgements}

\bibliographystyle{aa} % style aa.bst
\bibliography{references} % your references Yourfile.bib
\begin{appendix}
\section{Cosmological corner plots $\Lambda$CDM}
\label{app:lcdm}
\begin{figure}[!h]
\noindent\parbox{\textwidth}{
\includegraphics[width=\textwidth,height=22cm]{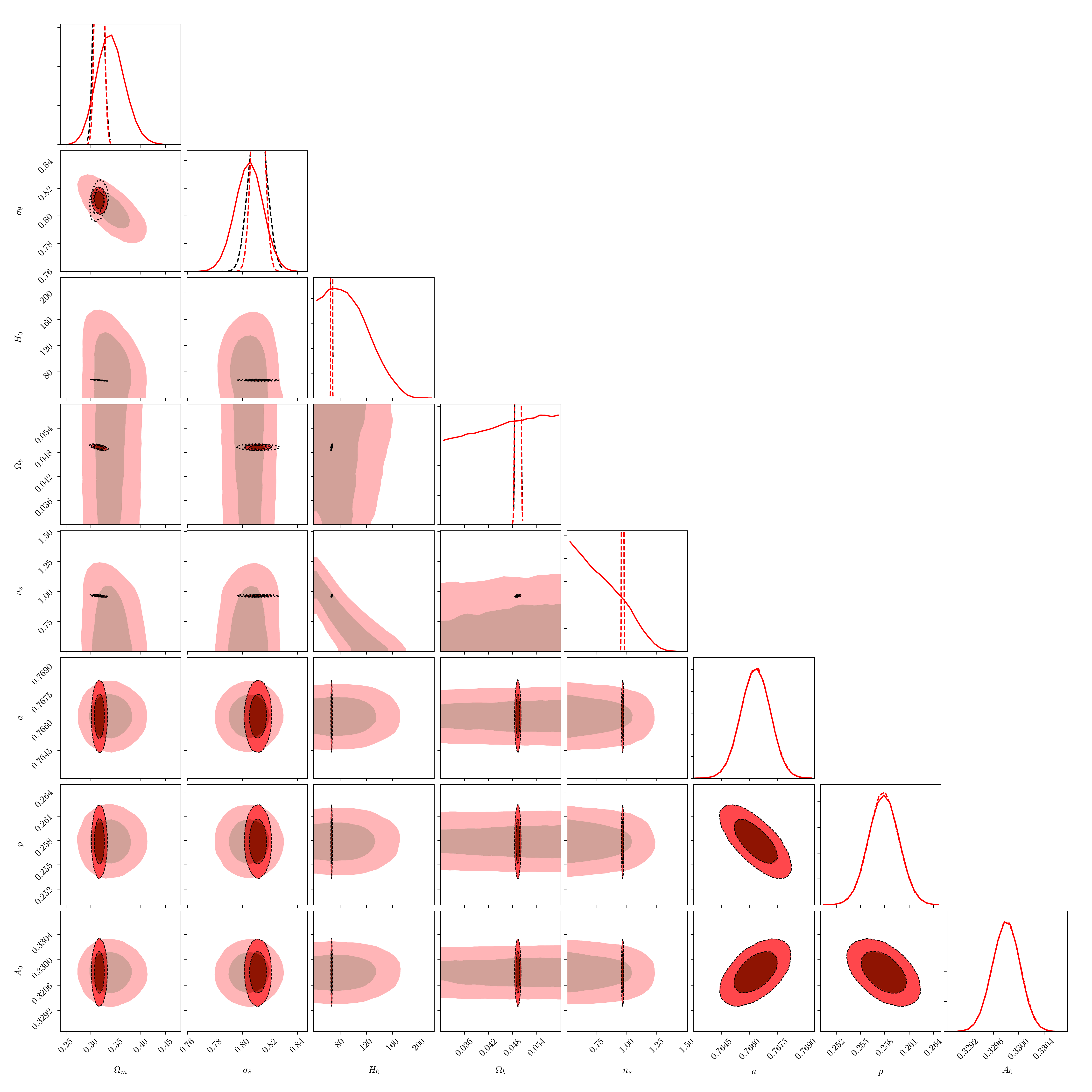}
}
\noindent\parbox{\textwidth}{\caption{Cosmological results (clusters only in light shading, Planck2018 with the black dotted line and clusters+Planck2018 in dark shading) for a a \Planck-like catalog of 513 objects in a $\Lambda$CDM cosmology}
}
\label{planck_lcdm}
\end{figure} 
\begin{figure*}[!h]
\includegraphics[width=\textwidth,height=23cm]{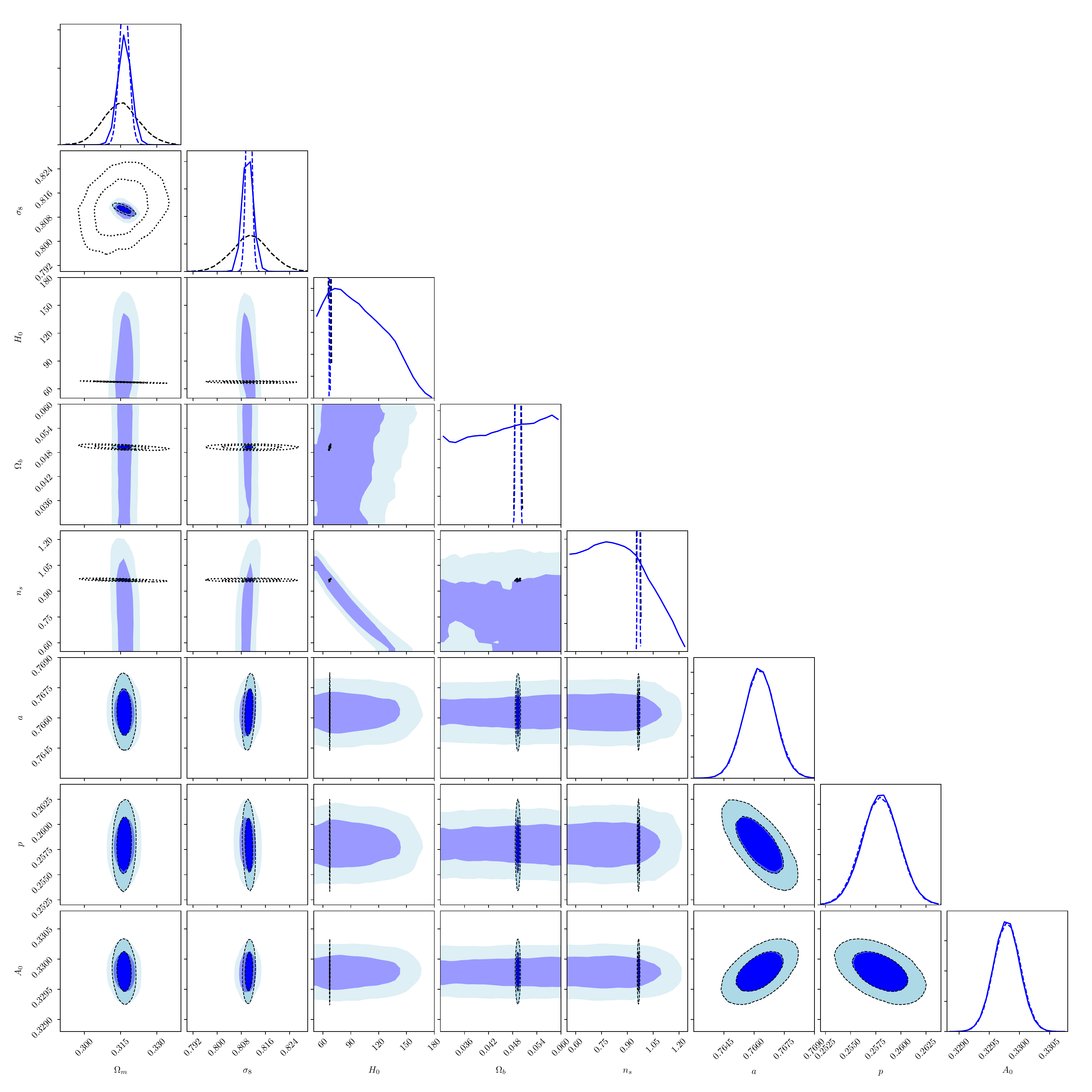}
\centering
\caption{Cosmological results (clusters only in light shading, Planck2018 with the black dotted line and clusters+Planck2018 in dark shading) for a \Euclid-like catalog of 66,252 objects in a $\Lambda$CDM cosmology}
\label{euclid_lcdm}
\end{figure*}
\begin{figure*}[!h]
\includegraphics[width=\textwidth,height=22cm]{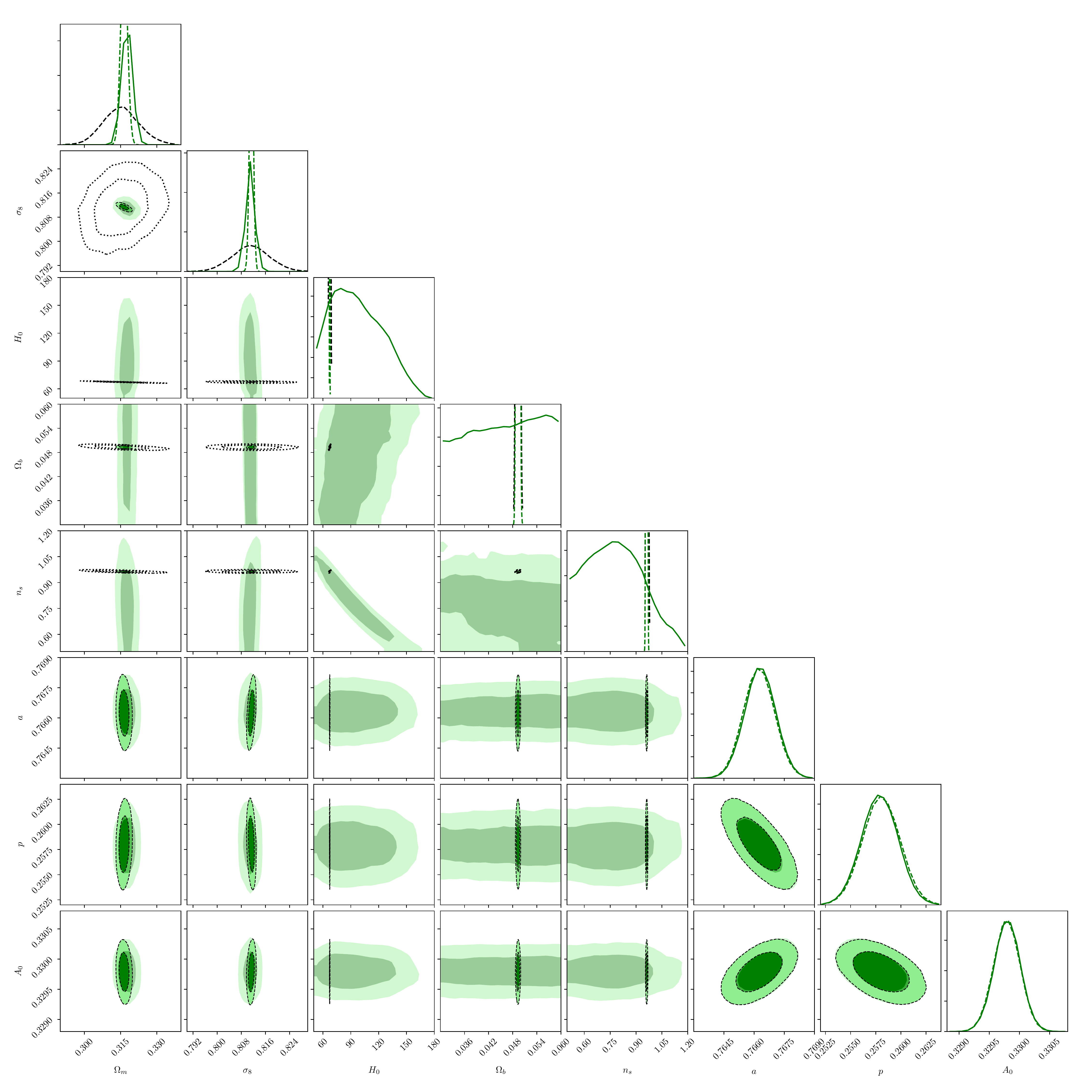}
\centering
\caption{Cosmological results (clusters only in light shading, Planck2018 with the black dotted line and clusters+Planck2018 in dark shading) for a Future catalog of 153,891 objects in a $\Lambda$CDM cosmology}
\label{futur_lcdm}
\end{figure*}
\newpage\phantom{skippage}
\newpage\phantom{skippage}
\newpage\phantom{skippage}
\section{Cosmological corner plots $w_0w_a$CDM}
\label{app:wowacdm}
\begin{figure}[!h]
\noindent\parbox{\textwidth}{
\includegraphics[width=\textwidth,height=21cm]{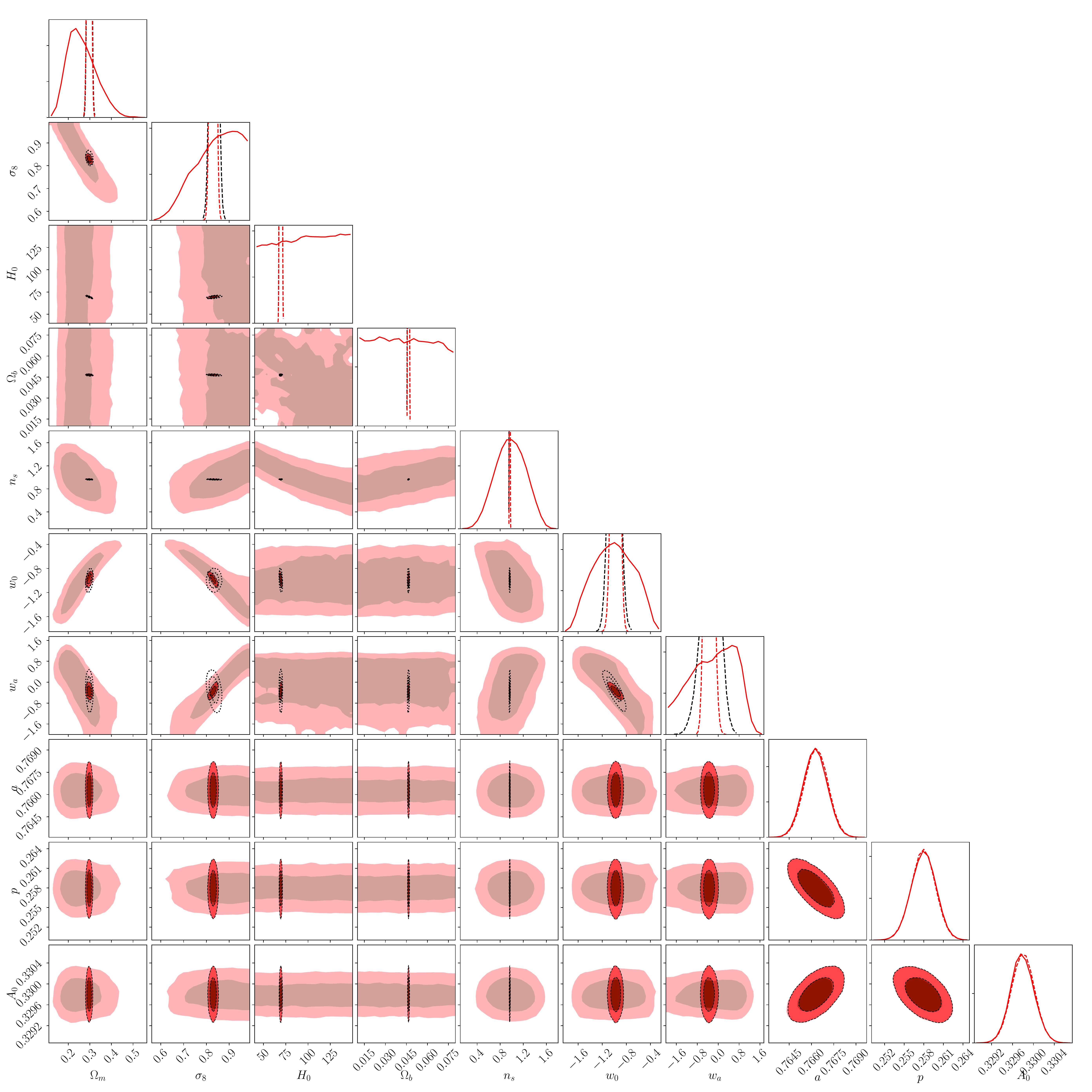}
}
\noindent\parbox{\textwidth}{\caption{Cosmological results (clusters only in light shading, Planck2018 with the black dotted line and clusters+Planck2018 in dark shading) for a \Planck-like catalog of 513 objects in a $w_0w_a$CDM cosmology}
}
\label{planck_wcdm}
\end{figure} 
\begin{figure*}[!h]
\includegraphics[width=\textwidth,height=21cm]{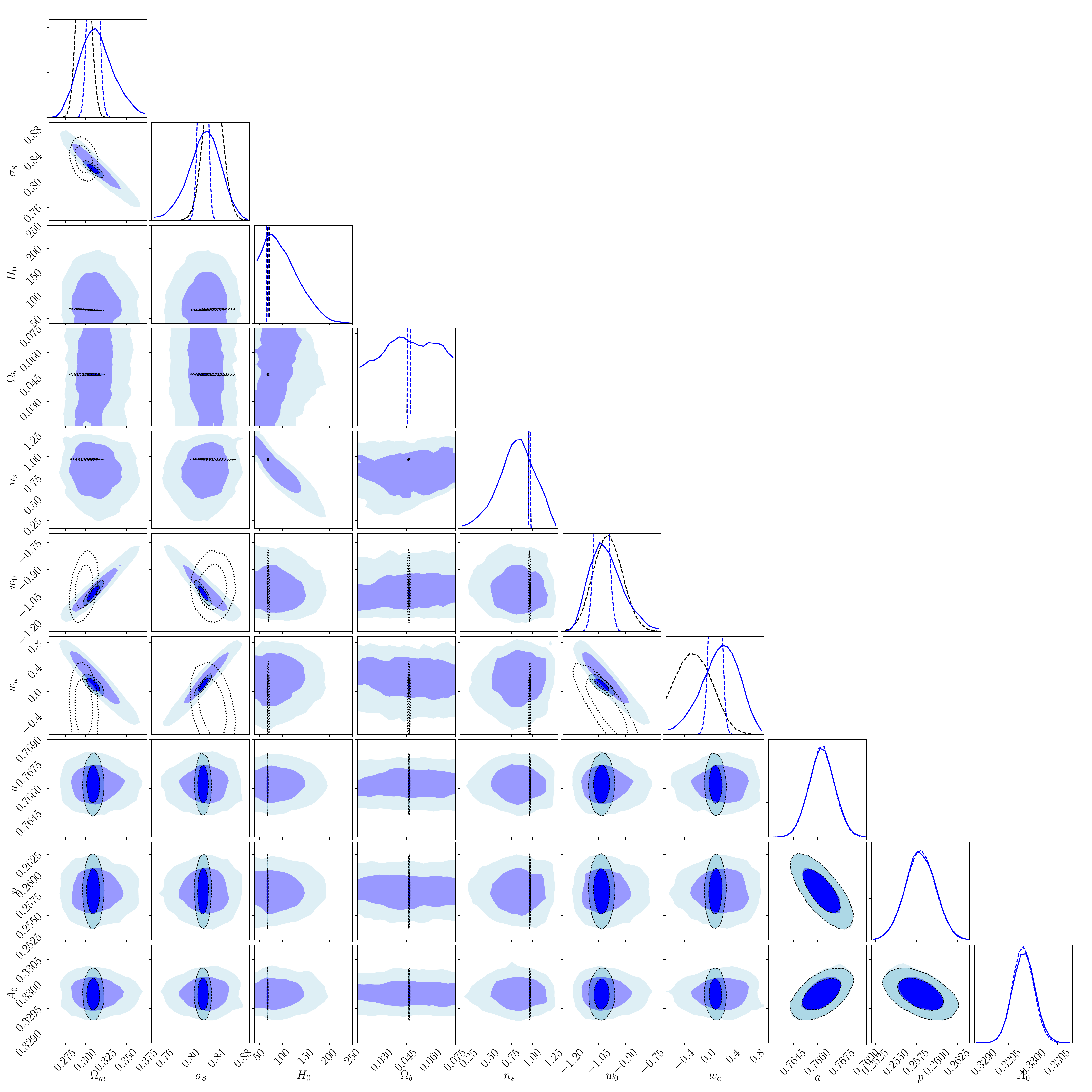}
\centering
\caption{Cosmological results (clusters only in light shading, Planck2018 with the black dotted line and clusters+Planck2018 in dark shading) for a \Euclid-like catalog of 66,252 objects in a $w_0w_a$CDM cosmology}
\label{euclid_wcdm}
\end{figure*}
\begin{figure*}[!h]
\includegraphics[width=\textwidth,height=22cm]{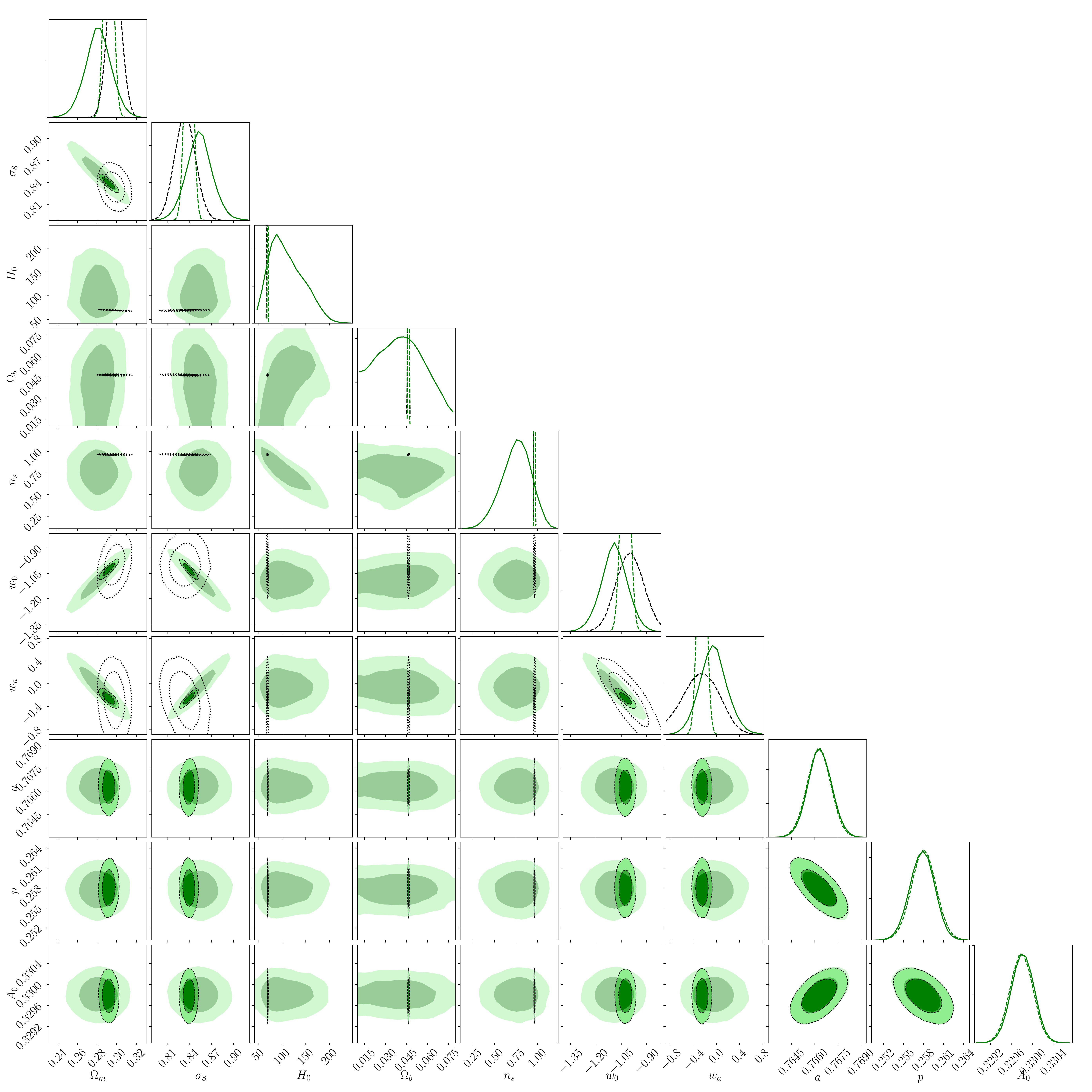}
\centering
\caption{Cosmological results (clusters only in light shading, Planck2018 with the black dotted line and clusters+Planck2018 in dark shading) for a Future catalog of 153,891 objects in a $w_0w_a$CDM cosmology}
\label{futur_wcdm}
\end{figure*}
\newpage\phantom{skippage}
\newpage\phantom{skippage}
\newpage\phantom{skippage}
\newpage\phantom{skippage}
\newpage\phantom{skippage}
\newpage\phantom{skippage}
%----------------------------------------
\section{Corner plots $\nu\mathrm{CDM}$}
\label{app:nucdm}
\begin{figure}[!h]
\noindent\parbox{\textwidth}{
\includegraphics[width=\textwidth,height=21cm]{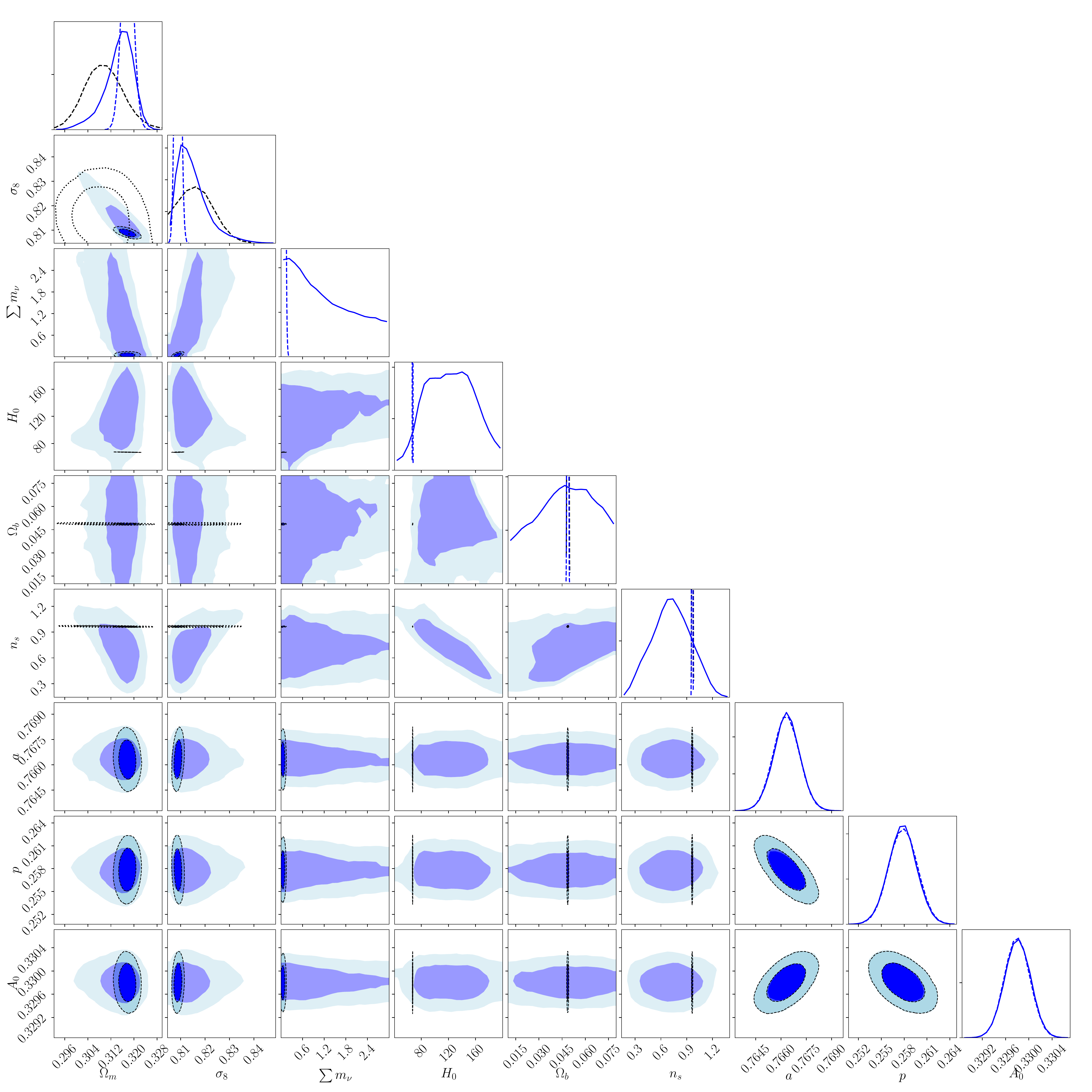}
}
\noindent\parbox{\textwidth}{\caption{Cosmological results (clusters only in light shading, Planck2018 with the black dotted line, clusters+Planck2018 in dark shading) for a \Euclid-like catalog of 65,994 objects in a $\nu$CDM cosmology}
}
\label{euclid_nucdm}
\end{figure} 
\begin{figure*}[!h]
\includegraphics[width=\textwidth,height=22cm]{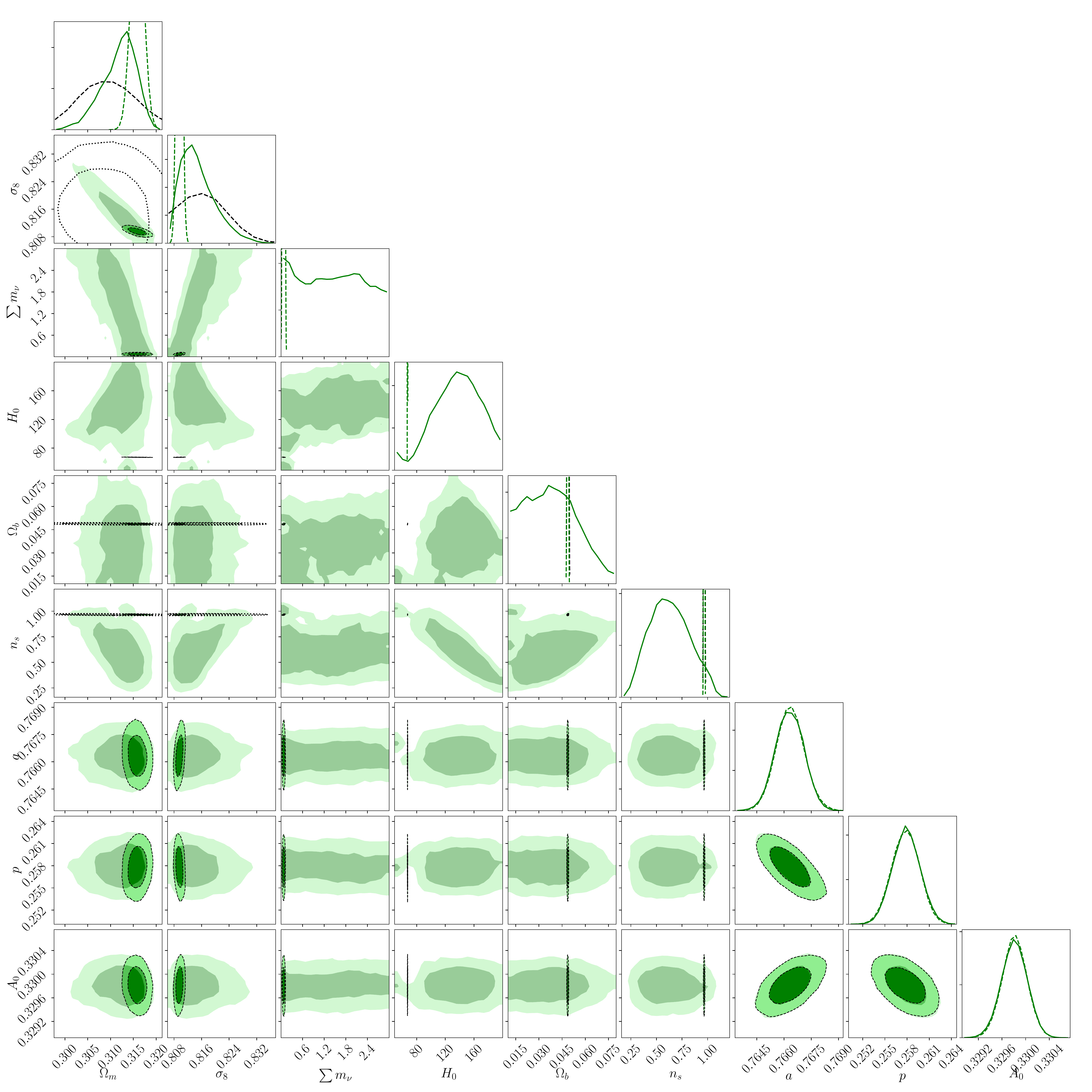}
\centering
\caption{Cosmological results (clusters only in light shading, Planck2018 with the black dotted line and clusters+Planck2018 in dark shading) for a catalog of 151,427 objects in a $\nu$CDM cosmology}
\label{future_nucdm}
\end{figure*}
\end{appendix}
\end{document}